\documentclass[submission,copyright,creativecommons]{eptcs}
\providecommand{\event}{WoC'15 post-proceedings} 

\usepackage{amsthm,amsmath,amssymb,bbm}
\usepackage{stmaryrd}
\usepackage{graphicx}

\newtheorem{theorem}{Theorem}

\newtheorem{proposition}[theorem]{Proposition}
\newtheorem{lemma}[theorem]{Lemma}

\newtheorem{definition}[theorem]{Definition }
\newtheorem{notation}[theorem]{Notation }

\newcommand{\qdatop}[1]{\genfrac{}{}{0pt}{0}{}{#1}}
\newcommand{\kid}[1]{\text{\it #1}}
\newcommand{\kwd}[1]{\textbf{\textup{#1}}}

\newcommand{\tmem}[1]{{\em #1\/}}
\newcommand{\tmop}[1]{\ensuremath{\text{#1}}}
\newcommand{\spc}{\;}

\newcommand{\longuparrow}{\uparrow}
\newcommand{\longdownarrow}{\downarrow}

\renewcommand{\mathbf}[1]{\text{\bf #1}}

\renewcommand{\prec}{\langle}
\renewcommand{\succ}{\rangle}

\title{A verified abstract machine for\\
functional coroutines}

\author{Tristan Crolard
\institute{CEDRIC, CNAM, Paris, France}
\email{tristan.crolard@cnam.fr}
}

\def\titlerunning{A verified abstract machine for functional coroutines}
\def\authorrunning{T. Crolard}

\begin{document}

\maketitle

\begin{abstract}\noindent
Functional coroutines are a restricted form of control mechanism, where each coroutine is represented with both
a continuation and an environment. This restriction was originally obtained by considering a constructive version of Parigot's classical natural deduction which is sound and complete for the Constant Domain logic. In this article, we present a refinement of de Groote's abstract machine for functional coroutines and we prove its correctness. Therefore, this abstract machine also provides a direct computational interpretation of the Constant Domain logic.

\end{abstract}

\section{Introduction}

The \textit{Constant Domain} logic (CD) is a well-known intermediate logic
due to Grzegorczyk {\cite{Grzegorczyk64}} which can be characterized as a
logic for Kripke frames with constant domains. Although CD is semantically
simpler than intuitionistic logic, its proof theory is quite difficult~: no
\textit{conventional} cut-free axiomatization is known
{\cite{Lopez-Escobar83}}, and it took more than three decades to prove that
the interpolation theorem does not hold either {\cite{Mints13}}. However, CD
is unavoidable when the object of study is \textit{duality in intuitionistic
logic}. Indeed, consider the following schema (called either D
{\cite{Grzegorczyk64}} or DIS {\cite{Rauszer80}}), where $x$ does not occur
free in $B$:
\[ \forall x (A \vee B) \vdash \left( \forall x \, A \right) \vee B \]
The dual of this schema is $\left( \exists x \, A \right) \wedge B \vdash
\exists x (A \wedge B)$ which is clearly valid in intuitionistic logic. Thus
bi-intuitionistic logic (also called Heyting-Brouwer logic {\cite{Rauszer80}}
or subtractive logic {\cite{Crolard01}}), which contains both intuitionistic
logic and dual intuitionistic logic, includes both schemas.

G$\ddot{\mathrm{o}}$rnemann proved that the addition of the DIS-schema to
intuitionistic predicate logic is sufficient to axiomatize CD
{\cite{Gornemann71}} (and also that the disjunction and existence properties
hold, so CD is still a constructive logic). Moreover, Rauszer proved that
bi-intuitionistic logic is conservative over CD {\cite{Rauszer80}} (Section~3,
p.~56), which means that the theorems of bi-intuitionistic logic with no
occurence of subtraction are exactly the theorems of CD. As a consequence, we
should expect at least the same difficulties with the proof-theoretical study
of bi-intuitionistic logic as with CD. In particular, if we want to understand
the computational content of bi-intuitionistic logic, it is certainly worth
spending some time on CD.

Although there is no conventional cut-free axiomatization of CD, there are
some non-conventional deduction systems which do enjoy cut elimination. The
first such system was defined by Kashima and Shimura
{\cite{Kashima91,Shimura94}} as a restriction of Gentzen's sequent calculus LK
based on dependency relations. Independently, we described {\cite{Crolard96}}
a similar restriction using Parigot's classical natural deduction
{\cite{Parigot94}} instead of LK. Another difference lies in the fact that our
restriction can also be formulated at the level of proof terms (terms of the
$\lambda \mu$-calculus in Parigot's system), independently of the typing
derivation. Such proof terms, which are terms of Parigot's $\lambda
\mu$-calculus, are called \textit{safe} in our calculus {\cite{Crolard04}}.
The intuition behind this terminology is presented informally in the
introduction of this article as follows:

\begin{quotation}
  ``[...] we observe that in the restricted $\lambda \mu$-calculus, even if
  continuations are no longer first-class objects, the ability of
  context-switching remains (in fact, this observation is easier to make in
  the framework of abstract state machines). However, a context is now a pair
  $\langle$\textit{environment}, \textit{continuation}$\rangle$. Note that
  such a pair is exactly what we expect as the context of a coroutine, since a
  coroutine should not access the local environment (the part of the
  environment which is not shared) of another coroutine. Consequently, we say
  that a $\lambda \mu$-term $t$ is \textit{safe with respect to coroutine
  contexts} (or just \textit{safe} for short) if no coroutines of $t$ access
  the local environment of another coroutine.''
\end{quotation}

In this paper, we provide some evidence to support this claim in the framework
of abstract state machines. As a starting point, we take an environment
machine for the $\lambda \mu$-calculus, which is defined and proved correct by
de Groote {\cite{deGroote98}} (a very similar machine was defined
independently by Streicher and Reus {\cite{Streicher98}}). Then we define a
new variant of this machine dedicated to the execution of safe terms which
works exactly as hinted above (let us call it the \textit{coroutine
machine}). Note that this modified machine is surprisingly simpler than what
we would expect form the negative proof-theoretic results. We actually obtain
a direct, meaningful, computational interpretation of the Constant Domain
logic, even though dependency relations were at the beginning only a complex
technical device.

As usual with environment machines, it is more convenient to encode variables
as de Bruijn indices (in particular for correctness proofs). Since safe
$\lambda \mu$-terms have different scoping rules than regular $\lambda
\mu$-terms, the translation into de Bruijn terms should yield different terms:
safe $\lambda \mu$-terms need to use \textit{local indices} to access the
local environment of the current coroutine, whereas arbitrary terms use the
usual \textit{global indices} to access the usual global environment.

As a consequence of this remark, we obtain a proof of correctness of the
coroutine machine which is two-fold. We first introduce an intermediate
machine which works with {\tmem{local indices, global environment and
indirection tables}}, then we show that this intermediate machine:
\begin{itemize}
  \item is simulated by de Groote's machine,
  
  \item is simulated by the coroutine machine.
\end{itemize}
We prove that both simulations are sound and complete, and as a consequence,
we obtain the correctness of the coroutine machine with respect to de Groote's
machine.

The plan of the paper is the following. In Section~\ref{safety}, we first
recall the notion of safety {\cite{Crolard96}}, and then we present a simpler
(but equivalent) definition of safety which is more convenient for correctness
proofs. In Section~\ref{machines}, we present our variant of de Groote's
machine and the coroutine machine. Finally, in Section~\ref{bisimulations}, we
detail the proof of correctness: we describe the intermediate machine and the
two simulations together with their properties (all the proofs were
mechanically checked with the Coq proof assistant, and the formalization is
available in the companion technical report {\cite{Crolard15}}).

\subsection{Related work}

\subsubsection*{Computational interpretation of classical logic}

Since Griffin's pioneering work {\cite{Griffin90}}, the extension of the
well-known formulas-as-types paradigm to classical logic has been widely
investigated for instance by \textrm{Murthy} {\cite{Murthy91b}},
\textrm{Barbanera} and \textrm{Berardi} {\cite{Berardi94b}},
\textrm{Rehof} and S{\o}rensen {\cite{Rehof94}}, \textrm{de Groote}
{\cite{deGroote01}}, and \textrm{Krivine} {\cite{Krivine94}}. We shall
consider here \textrm{Parigot}'s $\lambda \mu$-calculus mainly because it is
confluent and strongly normalizing in the second order framework
{\cite{Parigot94}}. Note that Parigot's original CND is a second-order logic,
in which $\vee, \wedge, \exists, \exists^2$ are definable from $\rightarrow,
\forall, \forall^2$. An extension of CND with primitive conjunction and
disjunction has also been investigated by Pym, Ritter and Wallen
{\cite{Pym00}} and de Groote {\cite{deGroote01}}.

The computational interpretation of classical logic is usually given by a
$\lambda$-calculus extended with some form of control (such as the famous
\textbf{call/cc} of Scheme or the catch/throw mechanism of Lisp) or similar
formulations of first-class continuation constructs. Continuations are used in
denotational semantics to describe control commands such as jumps
{\cite{Wadsworth00,Strachey74}}. They can also be used as a programming
technique to simulate backtracking and coroutines. For instance, first-class
continuations have been successfully used to implement Simula-like cooperative
coroutines in Scheme {\cite{Wand86}} or to provide simple and elegant
implementations of light-weight processes (or threads) {\cite{Dybvig89}}. This
approach has also been applied in Standard ML of New Jersey {\cite{Reppy95}}
using the typed counterpart of Scheme's \textbf{call/cc}
{\cite{MacQueen93}}. The key point in these implementations is that control
operators make it possible to switch between coroutine contexts, where the
context of a coroutine is encoded as its continuation.

\subsubsection*{Coroutines}

The concept of coroutine is usually attributed to Conway {\cite{Conway63}} who
introduced it to describe the interaction between a lexer and a parser inside
a compiler. They were also used by Knuth {\cite{Knuth97}} (Section~1.4.2,
p.~193) who saw them as a mechanism that generalizes subroutines (procedures
without parameters). Coroutines first appeared in a mainstream language in
Simula-67 {\cite{Dahl66}} and a formal framework for proving the correctness
of simple Simula programs containing coroutines has even been developed
{\cite{Clint73}}. Coroutine mechanisms were later introduced in several
programming languages, for instance in Modula-2 {\cite{Wirth80}}, and more
recently in the functional language Lua {\cite{Moura04a,Moura04}}.

Marlin's thesis {\cite{Marlin80}}, which is cited as a reference for
coroutines implementations {\cite{Moura04a}}, summarizes the characteristics
of a coroutine as follows:
\begin{enumerate}
  \item {\tmem{the values of data local to a coroutine persist between
  successive occasions on which control enters it (that is, between successive
  calls), and}}
  
  \item {\tmem{the execution of a coroutine is suspended as control leaves it,
  only to carry on where it left off when control re-enters the coroutine at
  some later stage.}}
\end{enumerate}
That is, a coroutine is a subroutine \textit{with a local state} which can
suspend and resume execution. This informal definition is of course not
sufficient to capture the various implementations that have been developed in
practice. To be more specific, the main differences between coroutine
mechanisms can be described as follows {\cite{Moura04a}}:
\begin{itemize}
  \item {\tmem{the control-transfer mechanism, which can provide symmetric or
  asymmetric coroutines.}}
  
  \item {\tmem{whether coroutines are provided in the language as first-class
  objects, which can be freely manipulated by the programmer, or as
  constrained constructs; }}
  
  \item {\tmem{whether a coroutine is a stackful construct, i.e., whether it
  is able to suspend its execution from within nested calls.}}
\end{itemize}
Symmetric coroutines generally offer a single control-transfer operation that
allows coroutines to pass control between them. Asymmetric control mechanisms,
sometimes called \textit{semi-coroutines} {\cite{Dahl72}}, rely on two
primitives for the transfer of control: the first to invoke a coroutine, the
second to pause and return control to the caller.

A well-known illustration of the third point above, called ``the same-fringe
problem'', is to determine whether two trees have exactly the same sequence of
leaves using two coroutines, where each coroutine recursively traverses a tree
and passes control to the other coroutine when it encounters a leaf. The
elegance of this algorithm lies in the fact that each coroutine uses its own
stack, which permits for two simple recursive tree traversals.

Although the first occurrence of the same-fringe problem in the litterature
seems indeed to be an illustration of a coroutine mechanism
{\cite{Prenner71}}, researchers did not agree on whether coroutines were
really required to solve this problem. In fact, several ``iterative''
solutions were then proposed for instance by Greussay {\cite{Greussay76}},
Anderson {\cite{Anderson76}} and McCarthy {\cite{McCarthy77}}. In fact, a
variety of inter-derivable solutions of this problem exist that do not solely
rely on coroutines {\cite{Biernacki06}}. However, since we are also interested
in program logics, it is relevant to quote McCarthy's conclusion about his own
solution: ``\textit{A program with only assignments and goto's may have the
most easily modified control structure. Of course, elegance, understandability
and a control logic admitting straightforward proofs of correctness are also
virtues}''.

More recently, Anton and Thiemann described a static type system for
first-class, stackful coroutines {\cite{Anton10}} that may be used in both,
symmetric and asymmetric ways. They followed Danvy's method {\cite{Danvy08}}
to derive definitional interpreters for several styles of coroutines from the
literature (starting from reduction semantics for Lua). This work is clearly
very close to our formalization, and it should help shed some light on these
mechanisms. However, we should keep in mind that logical deduction systems
come with their own constraints which might not be fully compatible with
existing programming paradigms: nobody knows to what extent what Griffin did
for continuations [22] can be done for coroutines.

{\paragraph{Remark.}{Asymmetric coroutines often correspond to the coroutines mechanism
made directly accessible to the programmer (as in Simula or Lua), sometimes as
a restricted form of \textit{generators} (as in C\#). On the other hand,
symmetric coroutines are generally chosen as a low-level mechanism used to
implement more advanced concurrency mechanisms (as in Modula). An other such
example is the Unix Standard {\cite{Open97}} where the recommended low-level
primitives for implementing lightweight processes (users threads) are
\textit{getcontext}, \textit{setcontext}, \textit{swapcontext} and
\textit{makecontext}. This is the terminology we have previously adopted for
our coroutines {\cite{Crolard04}}. However, since we are working in a purely
functional framework, we shall write ``functional coroutines'' to avoid any
confusion with other mechanisms.}}

\section{Dependency relations}\label{safety}

Parigot's original CND is a deduction system for the second-order classical
logic. Since we are mainly interested here in the computational content of
untyped terms, we shall simply recall the restriction in the propositional
framework corresponding to classical logic with the implication as only
connective (in Table\begin{table}[t!]
  \begin{eqnarray*}
    & x : \Gamma, A^x \vdash \Delta ; A & \\
    &  & \\
    & \dfrac{t : \Gamma, A^x \vdash \Delta ; B}{\lambda x.t : \Gamma \vdash
    \Delta ; A \rightarrow B}  (I_{\rightarrow}) \qquad \dfrac{t : \Gamma
    \vdash \Delta ; A \rightarrow B \hspace{1cm} u : \Gamma \vdash \Delta ;
    A}{t \; u : \Gamma \vdash \Delta ; B}  (E_{\rightarrow}) & \\
    &  & \\
    & \dfrac{t : \Gamma \vdash \Delta ; A}{\mathbf{throw} \text{ } \alpha
    \text{ } t : \Gamma \vdash \Delta, A^{\alpha} ; B}  (W_R) \qquad \qquad
    \dfrac{t : \Gamma \vdash \Delta, A^{\alpha} ; A}{\mathbf{catch} \text{ }
    \alpha \text{ } t : \Gamma \vdash \Delta ; A}  (C_R) & 
  \end{eqnarray*}
  
  \caption{Classical Natural Deduction\label{CND}}
\end{table}~\ref{CND}). We refer the reader to {\cite{Crolard96,Crolard04}}
for the full treatment of primitive conjunction, disjunction and quantifiers
(including the proof that the restricted system is sound and complete for CD).

{\paragraph{Remark.}{We actually work with a minor variant of the original $\lambda
\mu$-calculus, called the $\lambda_{\tmop{ct}}$-calculus, with a primitive
catch/throw mechanism {\cite{Crolard98a}}. These primitives are however easily
definable in the $\lambda \mu$-calculus as $\mathbf{catch} \; \alpha \text{ }
t \equiv \mu \alpha [\alpha] t$ and $\mathbf{throw} \; \alpha \text{ } t
\equiv \mu \delta [\alpha] t$ where $\delta$ is a name which does not occur in
$t$. }}

Since Parigot's CND is multiple-conclusioned sequent calculus, it is possible
to apply so-called \textit{Dragalin restriction} to obtain a sound and
complete system for CD. This restriction requires that the succedent of the
premise of the introduction rule for implication have only one formula:
\[ \frac{\Gamma, A \vdash B}{\Gamma \vdash \Delta, A \rightarrow B} \]
Unfortunately, the Dragalin restriction is not stable under proof reduction.
However, a weaker restriction which is stable under proof reduction, consists
in allowing multiple conclusions in the premise of this rule, with the proviso
that \textit{these other conclusions do not depend on $A$.} These
dependencies between occurrences of hypotheses and occurrences of conclusions
in a sequent are defined by induction on the derivation.

{\paragraph{Example.}{Consider a derived sequent $A, B^{}, C \vdash D, E, F, G$ with the
following dependencies:
\[ \text{\includegraphics[scale=1.1]{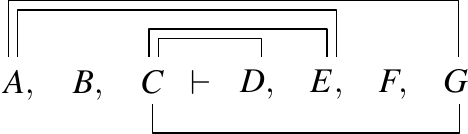}} \]
Using named hypotheses $A^x, B^y, C^z \vdash D, E, F, G$, this annotated
sequent may be represented as:
\[ A^x, B^y, C^z \vdash \{z\} : D, \{x, z\} : E, \{\} : F, \{x, z\} : G \]
Let us assume now that $t$ is the proof term corresponding to the above
derivation, i.e., we have derived in CND the following typing judgment:
\[ t : A^x, B^y, C^z \vdash D^{\alpha}, E^{\beta}, F^{\gamma} ; G \]
then we could also obtain the same dependencies directly from $t$, by
computing sets of variables used by the various coroutines (where $[]$ refers
to the distinguished conclusion), and we would get:
\begin{itemize}
  \item $\mathcal{S}_{\alpha} (t) = \{z\}$
  
  \item $\mathcal{S}_{\beta} (t) = \{x, z\}$
  
  \item $\mathcal{S}_{\gamma} (t) = \{\}$
  
  \item $\mathcal{S}_{[]} (t) = \{x, z\}$
\end{itemize}}}

{\paragraph{Remark.}{Deduction systems which rely on the Dragalin restriction usually do
not enjoy the cut elimination property: there are some derivable sequents for
which no cut-free proof exists. Pinto and Uustalu {\cite{Pinto09}} recently
presented such a counter-example (which is also mentionned by Gor{\'e} and
Postniece {\cite{Gore08}}) for Rauszer's sequent calculus for
bi-intuitionistic logic {\cite{Rauszer74}}. They show that there is no
cut-free proof of the following sequent in Rauszer calculus:
\[ p \vdash q, (r \rightarrow ((p - q) \wedge r)) \]
There is however a cut-free proof of this sequent in various extended sequent
calculi {\cite{Gore08,Pinto09,Pinto10}} and, as expected, there is also a
cut-free proof in the dependency-based system for bi-intuitionistic logic
{\cite{Crolard04}}:
\[ \dfrac{\dfrac{\text{\includegraphics[scale=1.1]{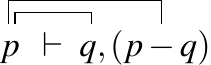}} \qquad
   \text{\includegraphics[scale=1.1]{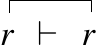}}}{\text{\includegraphics[scale=1.1]{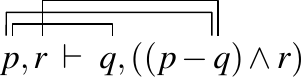}}}}{\text{\includegraphics[scale=1.1]{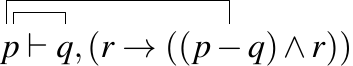}}}
\]
}}

The above presentation assumes that dependencies are explicitly displayed in
derivations. In fact, there is no need to actually annotate sequents with
dependency relations: the relevant information is already present inside the
proof term. Let us recall how these dependencies can be extracted
{\cite{Crolard04}}. In the following definition $\mathcal{S}_{\delta} (t)$
corresponds to the set of variables of $t$ which are \textit{used} by
coroutine $\delta$, whereas $\mathcal{S}_{[]} (t)$ corresponds to the set of
variables of $t$ which are \textit{used} by the ``current'' coroutine.

\begin{definition}
  Given a term $t$, for any free $\mu$-variable $\delta$ of $t$, the sets of
  variables $\mathcal{S}_{\delta} (v)$ and $\mathcal{S}_{[]} (u)$ are defined
  inductively as follows:
  \begin{itemize}
    \item $\begin{array}{l}
      \mathcal{S}_{[]} (x) = \{x\}\\
      \mathcal{S}_{\delta} (x) = \emptyset
    \end{array}$
    
    \
    
    \item $\begin{array}{l}
      \mathcal{S}_{[]} (\lambda x.u) =\mathcal{S}_{[]} (u) \backslash \{x\}\\
      \mathcal{S}_{\delta} (\lambda x.u) =\mathcal{S}_{\delta} (u) \backslash
      \{x\}
    \end{array}$
    
    \
    
    \item $\begin{array}{l}
      \mathcal{S}_{[]} (u \; v) =\mathcal{S}_{[]} (u) \cup \mathcal{S}_{[]}
      (v)\\
      \mathcal{S}_{\delta} (u \; v) =\mathcal{S}_{\delta} (u) \cup
      \mathcal{S}_{\delta} (v)
    \end{array}$
    
    \item $\begin{array}{l}
      \mathcal{S}_{[]} (\mathbf{catch} \; \alpha \text{ } u) =\mathcal{S}_{[]}
      (u) \cup \mathcal{S}_{\alpha} (u)\\
      \mathcal{S}_{\delta} (\mathbf{catch} \; \alpha \text{ } u)
      =\mathcal{S}_{\delta} (u)
    \end{array}$
    
    \item $\begin{array}{l}
      \mathcal{S}_{[]} (\mathbf{throw} \; \alpha \text{ } u) = \emptyset\\
      \mathcal{S}_{\alpha} (\mathbf{throw} \; \alpha \text{ } u)
      =\mathcal{S}_{\alpha} (u) \cup \mathcal{S}_{[]} (u)\\
      \mathcal{S}_{\delta} (\mathbf{throw} \; \alpha \text{ } u)
      =\mathcal{S}_{\delta} (u) \; \text{for any } \delta \neq \alpha
    \end{array}$
  \end{itemize}
\end{definition}

\begin{definition}
  A term $t$ is $\mathbf{safe}$ if and only if for any subterm of $t$ which
  has the form $\lambda x.u$, for any free $\mu$-variable $\delta$ of $u$, $x
  \notin \mathcal{S}_{\delta} (u)$ .
\end{definition}

{\paragraph{Example.}{The term $\lambda x. \mathbf{catch} \; \alpha \; \lambda y.
\mathbf{throw} \; \alpha \; x$ is safe, since $x$ was declared before
$\mathbf{catch} \; \alpha$ and $x$ is thus visible in $\mathbf{throw} \;
\alpha \; x$. On the other hand, $\lambda x. \mathbf{catch} \; \alpha \;
\lambda y. \mathbf{throw} \; \alpha \; y$ is not safe, because $y$ is not
visible in $\mathbf{throw} \; \alpha \; y$. More generally, for any $\alpha$,
a term of the form $\lambda y. \mathbf{throw} \; \alpha \; y$ is the
reification of $\alpha$ as a first-class continuation and such a term is never
safe. This can also be understood at the type level since the typing judgment
of such a term is the law of excluded middle $\vdash A^{\alpha} ; \neg A$.}}

{\paragraph{Remark.}{You can thus decide \textit{a posteriori} if a proof in CND is
valid in CD simply by checking if the (untyped) proof term is safe.}}

{\paragraph{Example.}{Here is a derivation of schema DIS in CND (using a primitive
disjunction):

{\small{\[ \dfrac{\dfrac{\dfrac{\dfrac{\dfrac{u : \forall x (A \vee B)^u
   \vdash \forall x (A \vee B)}{u : \forall x (A \vee B)^u \vdash A \vee B}
   \qquad \qdatop{A^a \vdash A} \qquad \dfrac{\dfrac{\dfrac{b : B^b \vdash
   B}{\mathbf{inr} \text{ } b : B^b \vdash (A \vee B)}}{\mathbf{inr} \text{ }
   b : B^b \vdash \forall x (A \vee B)}}{\mathbf{throw} \; \alpha \text{ }
   \left( \mathbf{inr} \text{ } b \right) : \forall x (A \vee B)^u \vdash
   \forall x (A \vee B)^{\alpha} ; A}}{\text{} \mathbf{case} \; \text{} u \;
   \mathbf{of} \hspace{1em} \mathbf{inl} \; a \rightarrow a \; \left| \;
   \mathbf{inr} \; b \rightarrow \mathbf{throw} \; \alpha \text{ } \left(
   \mathbf{inr} \text{ } b \right) : \forall x (A \vee B)^u \vdash \forall x
   (A \vee B)^{\alpha} ; A \right.}}{\text{} \mathbf{case} \; \text{} u \;
   \mathbf{of} \hspace{1em} \mathbf{inl} \; a \rightarrow a \; \left| \;
   \mathbf{inr} \; b \rightarrow \mathbf{throw} \; \alpha \text{ } \left(
   \mathbf{inr} \text{ } b \right) : \forall x (A \vee B)^u \vdash \forall x
   (A \vee B)^{\alpha} ; \forall xA \right.}}{\text{} \mathbf{inl} \; \left(
   \text{} \mathbf{case} \; \text{} u \; \mathbf{of} \hspace{1em} \mathbf{inl}
   \; a \rightarrow a \; \left| \; \mathbf{inr} \; b \rightarrow
   \mathbf{throw} \; \alpha \text{ } \left( \mathbf{inr} \text{ } b \right)
   \right) : \forall x (A \vee B)^u \vdash \forall x (A \vee B)^{\alpha} ;
   (\forall xA) \vee B \right.}}{\mathbf{catch} \; \alpha \text{ } \left(
   \text{} \mathbf{inl} \; \left( \text{} \mathbf{case} \; \text{} u \;
   \mathbf{of} \hspace{1em} \mathbf{inl} \; a \rightarrow a \; \left| \;
   \mathbf{inr} \; b \rightarrow \mathbf{throw} \; \alpha \text{ } \left(
   \mathbf{inr} \text{ } b \right) \right) \right) : \forall x (A \vee B)^u
   \vdash (\forall xA) \vee B \right.} \]
}}

It is possible to extend the definition of safety to the primitive disjunction
and then check that this proof term is safe {\cite{Crolard96,Crolard04}}. An
alternative consists in relying on the usual definition of disjunction in
Heyting arithmetic:
\[ A \vee B \equiv \exists x : \mathit{int}  (x = 0 \Rightarrow A \wedge x
   \neq 0 \Rightarrow B) \]
In this case, the proof term would be:
\[ \mathbf{catch} \; \alpha \text{ } \left( \mathbf{inl} \; \left( \mathbf{if}
   \; \text{} \pi_0 (u) = 0 \; \mathbf{then} \; \pi_1 (u) \; \mathbf{else} \;
   \mathbf{throw} \; \alpha \text{ } \left( \mathbf{inr} \text{ } \pi_2 (u)
   \right) \right) \right) \]
One can then check that this proof term is indeed safe (where integers, tuples
and projections can be encoded in the pure $\lambda$-calculus). Note however
that the standard second-order encoding of the disjunction in the pure
$\lambda$-calculus {\cite{Fortune83}} does not work for this derivation, since
the second branch of the \textbf{case} statement is not safe when encoded as
a $\lambda$-abstraction. }}

{\paragraph{Remark.}{As already noted in Troelstra's monograph {\cite{Troelstra73}}
(Section 1.11.3, p.~92), adding the DIS-schema to Heyting Arithmetic yields
classical arithmetic (Peano's Arithmetic). To be more specific, if atomic
formulas are decidable (which is the case in HA), one can prove using DIS that
\textit{any} formula is decidable. Fortunately, there is another way to
combine DIS with intuitionistic arithmetic which does not suffer from this
drawback. Indeed, Leivant introduced system \textbf{IT}$(\mathbbm{N})$
{\cite{Leivant02}} as an intuitionistic first-order theory where ``being a
natural number'' is expressed using a unary predicate and where quantifiers
over natural numbers need to be relativized. One can check using standard
Kripke semantics that \textbf{IT}$(\mathbbm{N})$ can be extended with
(non-relativized) DIS and that the resulting system is still conservative over
Heyting Arithmetic (when restricted to relativized formulas).}}

\subsection{Safety revisited}

In the conventional $\lambda$-calculus, there are two standard algorithms to
decide whether a term is closed: either you build inductively the set of free
variables (as a synthesized attribute) and then check that it is empty, or you
define a recursive function which takes as argument the set of declared
variables (as an inherited attribute), and checks that each variable has been
declared.

Similarly, for the $\lambda \mu$-calculus there are two ways of defining
safety:~the previous definition refined the standard notion of free variable
(by defining a set per free $\mu$-variable). In the following definition,
$\kid{\tmop{Safe}}$ takes as arguments the sets of visible variables for each
coroutine, and then decides for each variable, if the variable is visible in
the current coroutine. For a closed term, $\kid{\tmop{Safe}}$ is called with
$\mathcal{V}, \mathcal{V}_{\mu}$ both empty.

\begin{definition}
  The property $\kid{\tmop{Safe}}^{\mathcal{V}, \mathcal{V}_{\mu}} (t)$ is
  defined by induction on $t$ as follows:
  \begin{eqnarray*}
    \kid{\tmop{Safe}}^{\mathcal{V}, \mathcal{V}_{\mu}} (x) & = & x \in
    \mathcal{V}\\
    \kid{\tmop{Safe}}^{\mathcal{V}, \mathcal{V}_{\mu}} (t \; u) & = &
    \kid{\tmop{Safe}}^{\mathcal{V}, \mathcal{V}_{\mu}} (t) \wedge
    \kid{\tmop{Safe}}^{\mathcal{V}, \mathcal{V}_{\mu}} (u)\\
    \kid{\tmop{Safe}}^{\mathcal{V}, \mathcal{V}_{\mu}} (\lambda x.t) & = &
    \kid{\tmop{Safe}}^{(x : : \mathcal{V}), \mathcal{V}_{\mu}} (t)\\
    \kid{\tmop{Safe}}^{\mathcal{V}, \mathcal{V}_{\mu}} (\kwd{\tmop{catch}}
    \spc \alpha \spc t) & = & \kid{\tmop{Safe}}^{\mathcal{V}, (\alpha \mapsto
    \mathcal{V}; \mathcal{V}_{\mu})} (t)\\
    \kid{\tmop{Safe}}^{\mathcal{V}, \mathcal{V}_{\mu}} (\kwd{\tmop{throw}}
    \spc \alpha \spc t) & = & \kid{\tmop{Safe}}^{\mathcal{V}_{\mu} (\alpha),
    \mathcal{V}_{\mu}} (t)
  \end{eqnarray*}
  where:
  \begin{itemize}
    \item $\mathcal{V}$ is a list of variables
    
    \item $\mathcal{V}_{\mu}$ maps $\mu$-variables onto lists of variables
  \end{itemize}
\end{definition}

{\paragraph{Remark.}{This definition can also be seen as the reformulation, at the level
of proof terms, of the ``top-down'' definition of the restriction of CND from
{\cite{Brede09}} which was introduced in the framework of proof search.}}

As expected, we can show that the above two definitions of safety are
equivalent. More precisely, the following propositions are provable.

\begin{proposition}
  For any term $t$ and any mapping $\mathcal{V}_{\mu}$ such that $F V_{\mu}
  (t) \subseteq \kid{\tmop{dom}} (\mathcal{V}_{\mu})$, we have:
  $\kid{\tmop{Safe}}^{\mathcal{V}, \mathcal{V}_{\mu}} (t)$ implies
  $\mathcal{S}_{[]} (t) \subseteq \mathcal{V}$ and $\mathcal{S}_{\delta} (t)
  \subseteq \mathcal{V}_{\mu} (\delta)$ for any $\delta \in \kid{\tmop{dom}}
  (\mathcal{V}_{\mu})$ and $t$ is safe.
\end{proposition}

\begin{proposition}
  For any safe term $t$, for any set $\mathcal{V}$ such that $\mathcal{S}_{[]}
  (t) \subseteq \mathcal{V}$, for any mapping $\mathcal{V}_{\mu}$ such that $F
  V_{\mu} (t) \subseteq \kid{\tmop{dom}} (\mathcal{V}_{\mu})$ and
  $\mathcal{S}_{\delta} (t) \subseteq \mathcal{V}_{\mu} (\delta)$ for any
  $\delta \in F V_{\mu} (t)$, we have $\kid{\tmop{Safe}}^{\mathcal{V},
  \mathcal{V}_{\mu}} (t)$.
\end{proposition}

\section{Abstract machines}\label{machines}

In this section, we recall Groote's abstract machine for the $\lambda
\mu$-calculus {\cite{deGroote98}}, then we present the modified machine for
safe terms and we prove its correctness. But before we describe the abstract
machines, we need to move to a syntax using \textit{de Bruijn} indices, and
to adapt the definition of safety.

{\paragraph{Remark.}{Note that it is also possible to start with an abstract machine for
$\lambda \mu$-terms with names (instead of \textit{de Bruijn} indices) as
proposed by Streicher and Reus {\cite{Streicher98}}, and then define a variant
of this machine tailored for safe $\lambda \mu$-terms. The main advantage of
this approach would be to keep a single syntax (since safety is just a
predicate and no compilation is required). However, this apparent simplicity
is misleading: it actually more difficult to formally prove the correctness of
such a machine (since we have to deal with bound variables and
$\alpha$-conversion).}}

\subsection{Safe $\lambda_{\tmop{ct}}$-terms}

We rely on \textit{de Bruijn} indices for both kind of variables (the
regular variables and the $\mu$-variables) but they correspond to different
name spaces. Let us now call \textit{vector} a list of indices (natural
numbers), and \textit{table} a list of vectors. The definition of
\textit{Safe} given for named terms can be rephrased for \textit{de
Bruijn} terms as follows. For a closed term, $\kid{\tmop{Safe}}$ is called
with $\mathcal{I}, \; \mathcal{I}_{\mu}$ both empty and $n = 0$.

\begin{notation}
  We write ${}^{\backprime} g {}^{\backprime}$ for the term consisting only of
  the variable with index $g$ (this is just an explicit notation for the
  constructor that takes an index and builds a term). The rest of the syntax
  is standard for \textit{de Bruijn} terms. In particular, since
  $\mathbf{catch}$ is also a binder, it takes only a term as argument in
  \textit{de Bruijn} notation (this is similar to the
  $\lambda$-abstraction).
\end{notation}

\begin{definition}
  Given t: \textit{term}, $\mathcal{I}$: vector, $\mathcal{I}_{\mu}$: table
  and n: \textit{nat}, the property $\mathit{Safe}_n^{\mathcal{I},
  \mathcal{I}_{\mu}} (t)$ is defined inductively by the following rules:
  \[ \dfrac{n - g = k \qquad k \in \mathcal{I}}{\mathit{Safe}_n^{\mathcal{I},
     \mathcal{I}_{\mu}} ({}^{\backprime} g {}^{\backprime})} \]
  
  \[ \dfrac{\mathit{Safe}_n^{\mathcal{I}, \mathcal{I}_{\mu}} (t) \qquad
     \mathit{Safe}_n^{\mathcal{I}, \mathcal{I}_{\mu}}
     (u)}{\mathit{Safe}_n^{\mathcal{I}, \mathcal{I}_{\mu}} \left( t \; u
     \right)} \]
  
  \[ \dfrac{\mathit{Safe}_{n + 1}^{ (n + 1 : : \mathcal{I}),
     \mathcal{I}_{\mu}} (t)}{\mathit{Safe}_n^{\mathcal{I}, \mathcal{I}_{\mu}}
     \left( \lambda t \right)} \]
  
  \[ \dfrac{\mathit{Safe}_n^{\mathcal{I}, (\mathcal{I}: : \mathcal{I}_{\mu})}
     (t)}{\mathit{Safe}_n^{\mathcal{I}, \mathcal{I}_{\mu}} \left(
     \mathbf{catch} \; t \right)} \]
  
  \[ \dfrac{\mathcal{I}_{\mu} (\alpha)  = \mathcal{I}' \qquad
     \mathit{Safe}_n^{\mathcal{I}', \mathcal{I}_{\mu}}
     (t)}{\mathit{Safe}_n^{\mathcal{I}, \mathcal{I}_{\mu}} \left(
     \mathbf{throw} \; \alpha \; t \right)} \]
\end{definition}

{\paragraph{Remark.}{Note that $n$ is used to count occurrences of $\lambda$ from the
root of the term (seen as a tree), and such a number clearly uniquely
determines a $\lambda$ on a branch. Since they are the numbers stored in
$\mathcal{I}$, $\mathcal{I}_{\mu}$, a difference is computed for the base case
since \textit{de Bruijn} indices count $\lambda$ beginning with the leaf.}}

\subsection{From local indices to global indices}

In the framework of environment machines, de Bruijn indices are used to
represent variables in order to point directly to their denotation in the
environment (the closure which is bound to the variable). On the other hand,
the intuition behind the safety property is that for each continuation, there
is only a fragment of the environment which is visible (the local environment
of the coroutine).

In the modified machine, these indices should point to locations in the local
environment. Although the abstract syntaxes are isomorphic, it is more
convenient to introduce a new calculus (since indices in terms have different
semantics), where we can also rename {\textbf{catch/throw}} as
{\textbf{get-context/set-context}} (to be consistent with the new
semantics). Let us call $\lambda_{\mathbf{gs}}$-calculus the resulting
calculus, and let us now define formally the translation of
$\lambda_{\mathbf{gs}}$-terms onto (safe) $\lambda_{\mathbf{ct}}$-terms.

{\paragraph{Remark.}{In the Coq proof assistant, it is often more convenient to represent
partial functions as relations (since all functions are total in Coq we would
need option types to encode partial functions). In the sequel, we call
``functional'' or ``deterministic'' any relation which has been proved
functional.}}

\begin{definition}
  The functional relation $\downarrow_n^{\mathcal{I}, \mathcal{I}_{\mu}} (t) =
  t'$, with t: $\lambda_{\mathbf{gs}}$-\textit{term}, $t'$:
  $\lambda_{\mathbf{ct}}$-\textit{term}, $\mathcal{I}$: vector,
  $\mathcal{I}_{\mu}$: table and n: \textit{nat}, is defined inductively by
  the following rules:
  \[ \dfrac{n  - \mathcal{I} (l) = g}{\downarrow_n^{\mathcal{I},
     \mathcal{I}_{\mu}} ({}^{\backprime} l {}^{\backprime}) = ({}^{\backprime}
     g {}^{\backprime})} \]
  
  \[ \dfrac{\downarrow_n^{\mathcal{I}, \mathcal{I}_{\mu}} (t) = t' \qquad
     \downarrow_n^{\mathcal{I}, \mathcal{I}_{\mu}} (u) =
     u'}{\downarrow_n^{\mathcal{I}, \mathcal{I}_{\mu}} \left( t \; u \right) =
     \left( t' \; u' \right)} \]
  
  \[ \dfrac{\downarrow_{n + 1}^{ (n + 1 : : \mathcal{I}), \mathcal{I}_{\mu}}
     (t) = t'}{\downarrow_n^{\mathcal{I}, \mathcal{I}_{\mu}} \left( \lambda t
     \right) = \left( \lambda t' \right)} \]
  
  \[ \dfrac{\downarrow_n^{\mathcal{I}, (\mathcal{I}: : \mathcal{I}_{\mu})} (t)
     = t'}{\downarrow_n^{\mathcal{I}, \mathcal{I}_{\mu}} \left(
     \mathbf{get-context} \; t \right) = \left( \mathbf{catch} \; t' \right)}
  \]
  
  \[ \dfrac{\mathcal{I}_{\mu} (\alpha)  = \mathcal{I}' \qquad
     \downarrow_n^{\mathcal{I}', \mathcal{I}_{\mu}} (t) =
     t'}{\downarrow_n^{\mathcal{I}, \mathcal{I}_{\mu}} \left(
     \mathbf{set-context} \; \alpha \; t \right) = \left( \mathbf{throw} \;
     \alpha \; t' \right)} \]
\end{definition}

The shape of this definition is obviously very similar to the definition of
safety. Actually, we can prove that a $\lambda_{\mathbf{ct}}$-term is safe if
and only if it is the image of some $\lambda_{\mathbf{gs}}$-term by the
translation.

\begin{lemma}
  \label{safe-image}$\forall$ $\mathcal{I}$ $\mathcal{I}_{\mu}$ $n$ $t'$,
  $\mathit{Safe}_n^{\mathcal{I}, \mathcal{I}_{\mu}} (t') \hspace{1em}
  \leftrightarrow \hspace{1em} \exists t, \downarrow_n^{\mathcal{I},
  \mathcal{I}_{\mu}} (t) = t'$.
\end{lemma}

\subsection{The K$_{\tmop{ct}}$-machine for $\lambda_{\tmop{ct}}$-terms}

De Groote's machine {\cite{deGroote98}} is an extension of the well-known
Krivine's abstract machine (K-machine) which has already been studied
extensively in the literature {\cite{Danvy07}}. Moreover, this abstract
machine has also been derived mechanically from a contextual semantics of the
$\lambda \mu$-calculus with explicit substitutions using the method developed
by Biernacka and Danvy {\cite{Biernacka07}}, and it is thus correct by
construction. The K$_{\tmop{ct}}$-machine we describe below is a variant de
Groote's machine tailored for $\lambda_{\tmop{ct}}$-terms.

\subsubsection{Closure, environment, stack and state of the
K$_{\tmop{ct}}$-machine}

\begin{definition}
  A closure is an inductively defined tuple $[t, \mathcal{E},
  \mathcal{E}_{\mu}]$ with $t$ : \textit{term}, $\mathcal{E}$ : environment
  (where an environment is a closure list), $\mathcal{E}_{\mu}$ :
  \textit{stack} \textit{list} (where a stack is a closure list).
\end{definition}

\begin{definition}
  A state is defined as a tuple $\prec$ $t$, $\mathcal{E}$,
  $\mathcal{E}_{\mu}$, $\mathcal{S}$ $\succ$ where $[t, \mathcal{E},
  \mathcal{E}_{\mu}]$ is a closure and $\mathcal{S}$ is a stack.
\end{definition}

\subsubsection{Evaluation rules for the K$_{\tmop{ct}}$-machine}

\begin{definition}
  The deterministic transition relation $\sigma_1$ $\rightsquigarrow$
  $\sigma_2$, with $\sigma_1$, $\sigma_2$: \textit{state}, is defined
  inductively by the following rules (where $\mathcal{E}$(k) is the k-th
  closure in $\mathcal{E}$ and $\mathcal{E}_{\mu} (\alpha)$ is the $\alpha$-th
  stack in $\mathcal{E}_{\mu}$):
  \[ \dfrac{\mathcal{E} (k)  =  [t, \mathcal{E}', \mathcal{E}_{\mu}']}{\prec
     {}^{\backprime} k {}^{\backprime}, \mathcal{E}, \mathcal{E}_{\mu},
     \mathcal{S} \succ \rightsquigarrow \prec t, \mathcal{E}',
     \mathcal{E}_{\mu}', \mathcal{S} \succ} \]
  
  \[ \prec \left( t  u \right), \mathcal{E}, \mathcal{E}_{\mu}, \mathcal{S}
     \succ \rightsquigarrow \prec t, \mathcal{E}, \mathcal{E}_{\mu}, [u,
     \mathcal{E}, \mathcal{E}_{\mu}] : : \mathcal{S} \succ \]
  
  \[ \prec \lambda t, \mathcal{E}, \mathcal{E}_{\mu}, c : : \mathcal{S} \succ
     \rightsquigarrow \prec t, (c : : \mathcal{E}), \mathcal{E}_{\mu},
     \mathcal{S} \succ \]
  
  \[ \prec \mathbf{catch} \; t, \mathcal{E}, \mathcal{E}_{\mu}, \mathcal{S}
     \succ \rightsquigarrow \prec t, \mathcal{E}, (\mathcal{S}: :
     \mathcal{E}_{\mu}), \mathcal{S} \succ \]
  
  \[ \dfrac{\mathcal{E}_{\mu} (\alpha)  = \mathcal{S}'}{\prec \mathbf{throw}
     \; \alpha \; t, \mathcal{E}, \mathcal{E}_{\mu}, \mathcal{S} \succ
     \rightsquigarrow \prec t, \mathcal{E}, \mathcal{E}_{\mu}, \mathcal{S}'
     \succ} \]
\end{definition}

\subsection{The K$_{\tmop{gs}}$-machine for $\lambda_{\tmop{gs}}$-terms (with
local environments)}

As mentioned in the introduction, the modified abstract machine for
$\lambda_{\tmop{gs}}$-terms is a surprisingly simple variant of de Groote's
abstract machine, where a $\mu$-variable is mapped onto a pair
$\langle$\textit{environment}, \textit{continuation}$\rangle$ (a context)
and not only a continuation. As expected, the primitives
\textbf{get-context} and \textbf{set-context} respectively capture and
restore the local environment together with the continuation.

\subsubsection{Closure, environment, stack and state of the
K$_{\tmop{gs}}$-machine}

\begin{definition}
  A closure$_l$ is an inductively defined tuple $[t, \mathcal{L},
  \mathcal{L}_{\mu}, \mathcal{E}_{\mu}]$ where $t$: \textit{term},
  $\mathcal{L}$: environment$_l$ (where an environment$_l$ is a
  \textit{closure$_l$} list), $\mathcal{L}_{\mu}$: environment$_l$ list, and
  $\mathcal{E}_{\mu}$: stack$_l$ list (where a stack$_l$ is
  \textit{closure$_l$} list).
\end{definition}

{\paragraph{Remark.}{ For simplicity, we keep two distinct mappings in a closure,
$\mathcal{L}_{\mu}$ and $\mathcal{E}_{\mu}$, but they have the same domain,
which is the set of free $\mu$-variables. A $\mu$-variable is then mapped onto
a pair $\langle$\textit{environment}, \textit{continuation}$\rangle$ as
expected: the environment is obtained from $\mathcal{L}_{\mu}$ and the
continuation is obtained from $\mathcal{E}_{\mu}$.}}

\begin{definition}
  A state$_l$ is defined as a tuple $\prec$ $t, \mathcal{L},
  \mathcal{L}_{\mu}, \mathcal{E}_{\mu}, \mathcal{S}$ $\succ$ where $[t,
  \mathcal{L}, \mathcal{L}_{\mu}, \mathcal{E}_{\mu}]$ is a closure$_l$ and
  $\mathcal{S}$ is a \textit{stack$_l$}.
\end{definition}

\subsubsection{Evaluation rules for the K$_{\tmop{gs}}$-machine}

\begin{definition}
  The deterministic transition relation $\sigma_1$ $\rightsquigarrow^l$
  $\sigma_2$, with $\sigma_1$, $\sigma_2$: \textit{state$_l$}, is defined
  inductively by the following rules:
  \[ \dfrac{\mathcal{L} (k)  =  [t, \mathcal{L}', \mathcal{L}_{\mu}',
     \mathcal{E}_{\mu}']}{\prec {}^{\backprime} k {}^{\backprime},
     \mathcal{L}, \mathcal{L}_{\mu}, \mathcal{E}_{\mu}, \mathcal{S} \succ
     \rightsquigarrow^l \prec t, \mathcal{L}', \mathcal{L}_{\mu}',
     \mathcal{E}_{\mu}', \mathcal{S} \succ} \]
  
  \[ \prec \left( t  u \right), \mathcal{L}, \mathcal{L}_{\mu},
     \mathcal{E}_{\mu}, \mathcal{S} \succ \rightsquigarrow^l \prec t,
     \mathcal{L}, \mathcal{L}_{\mu}, \mathcal{E}_{\mu}, [u, \mathcal{L},
     \mathcal{L}_{\mu}, \mathcal{E}_{\mu}] : : \mathcal{S} \succ \]
  
  \[ \prec \lambda t, \mathcal{L}, \mathcal{L}_{\mu}, \mathcal{E}_{\mu}, c : :
     \mathcal{S'} \succ \rightsquigarrow^l \prec t, (c : : \mathcal{L}),
     \mathcal{L}_{\mu}, \mathcal{E}_{\mu}, \mathcal{S'} \succ \]
  
  \[ \prec \mathbf{get-context} \; t, \mathcal{L}, \mathcal{L}_{\mu},
     \mathcal{E}_{\mu}, \mathcal{S} \succ \rightsquigarrow^l \prec t,
     \mathcal{L}, (\mathcal{L}: : \mathcal{L}_{\mu}), (\mathcal{S}: :
     \mathcal{E}_{\mu}), \mathcal{S} \succ  \]
  
  \[ \dfrac{\mathcal{L}_{\mu} (\alpha)  = \mathcal{L}' \qquad
     \mathcal{E}_{\mu} (\alpha)  = \mathcal{S}'}{\prec \mathbf{set-context} \;
     \alpha \; t, \mathcal{L}, \mathcal{L}_{\mu}, \mathcal{E}_{\mu},
     \mathcal{S} \succ \rightsquigarrow^l \prec t, \mathcal{L}',
     \mathcal{L}_{\mu}, \mathcal{E}_{\mu}, \mathcal{S}' \succ } \]
\end{definition}

\section{Bisimulations}\label{bisimulations}

We first introduce the intermediate machine for $\lambda_{\tmop{gs}}$-terms,
called the K$^{\tmop{it}}_{\tmop{gs}}$-machine, and we define two simulations
$\left( \mathsf{-} \right)^{\star}$ and $\left( \mathsf{-} \right)^{\diamond}$
showing that this K$^{\tmop{it}}_{\tmop{gs}}$-machine is simulated by both the
K$_{\tmop{ct}}$-machine and the K$_{\tmop{gs}}$-machine. Moreover, we shall
prove that both simulations are sound and complete.

\subsection{The K$^{\tmop{it}}_{\tmop{gs}}$-machine for
$\lambda_{\tmop{gs}}$-terms (with indirection tables)}

This intermediate machine for $\lambda_{\tmop{gs}}$-terms works {\tmem{with
local indices, global environment and indirection tables}}. The indirection
tables are exactly the same as for the static translation of
$\lambda_{\tmop{gs}}$-terms to safe $\lambda_{\tmop{ct}}$-terms. However, the
translation is now performed at runtime. The lock-step simulation $\left(
\mathsf{-} \right)^{\star}$ shows that translating during evaluation is indeed
equivalent to evaluating the translated term. The lock-step simulation $\left(
\mathsf{-} \right)^{\diamond}$ shows that we can ``flatten away'' the
indirection tables and the global environment, and work only with local
environments.

\subsubsection{Closure, environment, stack and state of the
K$^{\tmop{it}}_{\tmop{gs}}$-machine}

\begin{definition}
  A closure$_i$ is an inductively defined tuple $[t, n, \mathcal{I},
  \mathcal{I}_{\mu}, \mathcal{E}, \mathcal{E}_{\mu}]$, with $t$ :
  \textit{term}, $n$ : \textit{nat}, $\mathcal{I}$ : \textit{vector},
  $\mathcal{I}_{\mu}$ : \textit{table}, $\mathcal{E}$ :
  \textit{environment$_i$} (where an \textit{environment$_i$} is a
  \textit{closure$_i$} list), $\mathcal{E}_{\mu}$ : \textit{stack$_i$}
  list (where a \textit{stack$_i$} is a \textit{closure$_i$} list).
\end{definition}

\begin{definition}
  A state$_i$ is defined as a tuple $\prec$ $t$, $n$, $\mathcal{I}$,
  $\mathcal{I}_{\mu}$, $\mathcal{E}$, $\mathcal{E}_{\mu}$, $\mathcal{S}$
  $\succ$ where $[t, n, \mathcal{I}, \mathcal{I}_{\mu}, \mathcal{E},
  \mathcal{E}_{\mu}]$ is a closure$_i$ and $\mathcal{S}$ is a
  \textit{stack$_i$}.
\end{definition}

\subsubsection{Evaluation rules for the K$^{\tmop{it}}_{\tmop{gs}}$-machine}

\begin{definition}
  The deterministic transition relation $\sigma_1$ $\rightsquigarrow^i$
  $\sigma_2$, with $\sigma_1$, $\sigma_2$: \textit{state$_i$}, is defined
  inductively by the following rules:

  \[ \dfrac{n  - \mathcal{I} ( l )  = g \qquad \mathcal{E}
     (g)  =  [t, n', \mathcal{I}', \mathcal{I}_{\mu}', \mathcal{E}',
     \mathcal{E}_{\mu}']}{\prec {}^{\backprime} l {}^{\backprime}, n,
     \mathcal{I}, \mathcal{I}_{\mu}, \mathcal{E}, \mathcal{E}_{\mu},
     \mathcal{S} \succ \rightsquigarrow^i \prec t, n', \mathcal{I}',
     \mathcal{I}_{\mu}', \mathcal{E}', \mathcal{E}_{\mu}', \mathcal{S} \succ}
  \]
  
  \[ \prec \left( t  u \right), n, \mathcal{I}, \mathcal{I}_{\mu},
     \mathcal{E}, \mathcal{E}_{\mu}, \mathcal{S} \succ \rightsquigarrow^i
     \prec t, n, \mathcal{I}, \mathcal{I}_{\mu}, \mathcal{E},
     \mathcal{E}_{\mu}, [u, n, \mathcal{I}, \mathcal{I}_{\mu}, \mathcal{E},
     \mathcal{E}_{\mu}] : : \mathcal{S} \succ \]
  
  \[ \prec \lambda t, n, \mathcal{I}, \mathcal{I}_{\mu}, \mathcal{E},
     \mathcal{E}_{\mu}, c : : \mathcal{S'} \succ \rightsquigarrow^i \prec t,
     (S n), ((S n) : : \mathcal{I}), \mathcal{I}_{\mu}, c : : \mathcal{E},
     \mathcal{E}_{\mu}, \mathcal{S'} \succ \]
  
  \[ \prec \mathbf{get-context} \; t, n, \mathcal{I}, \mathcal{I}_{\mu},
     \mathcal{E}, \mathcal{E}_{\mu}, \mathcal{S} \succ \rightsquigarrow^i
     \prec t, n, \mathcal{I}, (\mathcal{I}: : \mathcal{I}_{\mu}), \mathcal{E},
     (\mathcal{S}: : \mathcal{E}_{\mu}), \mathcal{S} \succ  \]
  
  \[ \dfrac{\mathcal{I}_{\mu} (\alpha)  = \mathcal{I}' \qquad
     \mathcal{E}_{\mu} (\alpha)  = \mathcal{S}'}{\prec \mathbf{set-context} \;
     \alpha \; t, n, \mathcal{I}, \mathcal{I}_{\mu}, \mathcal{E},
     \mathcal{E}_{\mu}, \mathcal{S} \succ \rightsquigarrow^i \prec t, n,
     \mathcal{I}', \mathcal{I}_{\mu}, \mathcal{E}, \mathcal{E}_{\mu},
     \mathcal{S}' \succ } \]
\end{definition}

\subsection{Lock-step simulation $\left( \mathsf{-} \right)^{\star}$ of the
K$^{\tmop{it}}_{\tmop{gs}}$-machine by the K$_{\tmop{ct}}$-machine}

\begin{definition}
  The functional relation $c^{\star} =_c c'$, with $c$:
  \textit{closure$_i$}, \ $c'$: \textit{closure}, is defined by the
  following rule:
  \[ \dfrac{\downarrow_n^{\mathcal{I}, \mathcal{I}_{\mu}} (t) = u \qquad
     \mathcal{E}^{\star} =_e \mathcal{E}' \qquad \mathcal{E}_{\mu}^{\star} =_k
     \mathcal{E}_{\mu}'}{[t, n, \mathcal{I}, \mathcal{I}_{\mu}, \mathcal{E},
     \mathcal{E}_{\mu}]^{\star} =_c  [u, \mathcal{E}', \mathcal{E}_{\mu}']} \]
  where $\mathcal{E}^{\star}$ and $\mathcal{E}_{\mu}^{\star}$ are defined by
  element-wise application of $_{} ^{\star}$.
\end{definition}

\begin{definition}
  The functional relation $\sigma^{\star} =_{\sigma} \sigma'$, with $\sigma$:
  \textit{state$_i$}, \ $\sigma'$: \textit{state}, is defined by the
  following rule:
  \[ \dfrac{[t, n, \mathcal{I}, \mathcal{I}_{\mu}, \mathcal{E},
     \mathcal{E}_{\mu}]^{\star} =_c  [u, \mathcal{E}', \mathcal{E}_{\mu}']
     \qquad \mathcal{S}^{\star} =_s \mathcal{S}'}{\prec t, n, \mathcal{I},
     \mathcal{I}_{\mu}, \mathcal{E}, \mathcal{E}_{\mu}, \mathcal{S}
     \succ^{\star} =_{\sigma} \prec u, \mathcal{E}', \mathcal{E}_{\mu}',
     \mathcal{S}' \succ} \]
  where $\mathcal{S}^{\star}$ is defined by element-wise application of $_{}
  ^{\star}$.
\end{definition}

\subsubsection{Soundness of simulation $\left( \mathsf{-} \right)^{\star}$}

The following theorem states that the K$^{\tmop{it}}_{\tmop{gs}}$-machine is
sound with respect to the K$_{\tmop{ct}}$-machine.

\begin{theorem}
  $\forall$ $\sigma_1$ $\sigma_2$ $\sigma_1'$, $\sigma_1 \rightsquigarrow^i
  \sigma_2 \; \rightarrow \; \sigma_1^{\star} =_{\sigma} \sigma_1' \;
  \rightarrow \; \exists \sigma_2', \; \sigma_1' \rightsquigarrow \sigma_2' \;
  \wedge \; \sigma_2^{\star} =_{\sigma} \sigma_2'$.
\end{theorem}

\subsubsection{Completeness of simulation $\left( \mathsf{-} \right)^{\star}$}

The following theorem states that the K$^{\tmop{it}}_{\tmop{gs}}$-machine is
complete with respect to the K$_{\tmop{ct}}$-machine.

\begin{theorem}
  $\forall$ $\sigma_1'$ $\sigma_2'$ $\sigma_1$, $\sigma_1' \rightsquigarrow
  \sigma_2' \; \rightarrow \; \sigma_1^{\star} =_{\sigma} \sigma_1' \;
  \rightarrow \; \exists \sigma_2, \; \sigma_1 \rightsquigarrow^i \sigma_2 \;
  \wedge \; \sigma_2^{\star} =_{\sigma} \sigma_2'$.
\end{theorem}

{\paragraph{Remark.}{Since simulation $\left( \mathsf{-} \right)^{\star}$ is sound and
complete, and since both the K$^{\tmop{it}}_{\tmop{gs}}$-machine and
K$_{\tmop{gs}}$-machine evaluate the same $\lambda_{\tmop{gs}}$-terms,
simulation $\left( \mathsf{-} \right)^{\star}$ is actually a bi-simulation.}}

\subsection{Lock-step simulation $\left( \mathsf{-} \right)^{\diamond}$ of the
K$^{\tmop{it}}_{\tmop{gs}}$-machine by the K$_{\tmop{gs}}$-machine}

\begin{definition}
  The functional relation $c^{\diamond} =_k c'$ with $c$:
  \textit{closure$_i$}, \ $c'$: \textit{closure$_l$}, is defined by the
  following rule:
  \[ \dfrac{\mathit{flatten} \; n \; \mathcal{E} \; \mathcal{I} \; = \;
     \mathcal{L} \qquad \mathit{map} \; \left( \mathit{flatten} \; n \;
     \mathcal{E} \right) \; \mathcal{I}_{\mu}  \; = \; \mathcal{L}_{\mu}
     \qquad \mathcal{E}_{\mu}^{\diamond} =_k \mathcal{E}_{\mu}'}{[t, n,
     \mathcal{I}, \mathcal{I}_{\mu}, \mathcal{E},
     \mathcal{E}_{\mu}]^{\diamond} =_c  [t, \mathcal{L}, \mathcal{L}_{\mu},
     \mathcal{E}_{\mu}']} \]
  where $\mathcal{S}^{\diamond}$ and $\mathcal{E}_{\mu}^{\diamond}$ are
  defined by element-wise application of $_{} ^{\diamond}$, and flatten is a
  functional relation inductively defined by the following rules:
  \[ \mathit{flatten} \; n \; \mathcal{E} \; \mathit{nil} \; = \; \mathit{nil}
     \qquad \dfrac{\mathcal{E} (n  - k)  = c \qquad c^{\diamond} =_c c' \qquad
     \mathit{flatten} \; n \; \mathcal{E} \; \mathcal{I} \; = \;
     \mathcal{L}}{\mathit{flatten} \; n \; \mathcal{E} \; (k : : \mathcal{I})
     \; = \; (c' : : \mathcal{L})} \]
\end{definition}

\begin{definition}
  The functional relation $\sigma^{\diamond} =_{\sigma} \sigma'$, with
  $\sigma$ : \textit{state$_i$}, \ $\sigma'$: \textit{state$_l$}, is
  defined by the following rule:
  \[ \dfrac{[t, n, \mathcal{I}, \mathcal{I}_{\mu}, \mathcal{E},
     \mathcal{E}_{\mu}]^{\diamond} =_c  [u, \mathcal{L}, \mathcal{L}_{\mu},
     \mathcal{E}_{\mu}'] \qquad \mathcal{S}^{\diamond} =_s \mathcal{S}'}{\prec
     t, n, \mathcal{I}, \mathcal{I}_{\mu}, \mathcal{E}, \mathcal{E}_{\mu},
     \mathcal{S} \succ^{\diamond} =_{\sigma} \prec u, \mathcal{L},
     \mathcal{L}_{\mu}, \mathcal{E}_{\mu}', \mathcal{S}' \succ} \]
\end{definition}

\subsubsection{Soundness of simulation $\left( \mathsf{-} \right)^{\diamond}$}

The following theorem states that the K$^{\tmop{it}}_{\tmop{gs}}$-machine is
sound with respect to the K$_{\tmop{gs}}$-machine.

\begin{theorem}
  $\forall$ $\sigma_1$ $\sigma_2$ $\sigma_1'$, $\sigma_1 \rightsquigarrow^i
  \sigma_2 \; \rightarrow \; \sigma_1^{\diamond} =_{\sigma} \sigma_1' \;
  \rightarrow \; \exists \sigma_2', \; \sigma_1' \rightsquigarrow^l \sigma_2'
  \; \wedge \; \sigma_2^{\diamond} =_{\sigma} \sigma_2'$.
\end{theorem}

\subsubsection{Completeness of simulation $\left( \mathsf{-}
\right)^{\diamond}$}

The following theorem states that the K$^{\tmop{it}}_{\tmop{gs}}$-machine is
complete with respect to the K$_{\tmop{gs}}$-machine.

\begin{theorem}
  $\forall$ $\sigma_1'$ $\sigma_2'$ $\sigma_1$, $\sigma_1' \rightsquigarrow^l
  \sigma_2' \; \rightarrow \; \sigma_1^{\diamond} =_{\sigma} \sigma_1' \;
  \rightarrow \; \exists \sigma_2, \; \sigma_1 \rightsquigarrow^i \sigma_2 \;
  \wedge \; \sigma_2^{\diamond} =_{\sigma} \sigma_2'$.
\end{theorem}

{\paragraph{Remark.}{Since simulation $\left( \mathsf{-} \right)^{\diamond}$ is sound and
complete, $\left( \mathsf{-} \right)^{\diamond}$ is a bi-simulation when the
initial states of the K$_{\tmop{ct}}$-machine are restricted to safe
$\lambda_{\tmop{ct}}$-terms (by Lemma~\ref{safe-image}). However, the
K$_{\tmop{ct}}$-machine can also evaluate arbitrary
$\lambda_{\tmop{ct}}$-terms.}}

\subsection{Lock-step simulation of the K$_{\tmop{gs}}$-machine by the
K$_{\tmop{ct}}$-machine}

We have proved that the K$^{\tmop{it}}_{\tmop{gs}}$-machine is simulated by
both the K$_{\tmop{ct}}$-machine and the K$_{\tmop{gs}}$-machine (and that
both simulations are sound and complete). These properties are illustrated by
the following diagram:
\[ \begin{array}{lccccccccc}
     \text{K$_{\tmop{ct}}$-machine } & \sigma^{\star}_0 & \rightsquigarrow &
     \cdot \cdot \cdot & \rightsquigarrow & \sigma^{\star}_n &
     \rightsquigarrow & \sigma^{\star}_{n + 1} & \rightsquigarrow & \cdot
     \cdot \cdot\\
     & \longuparrow \star &  &  &  & \longuparrow \star &  & \longuparrow
     \star &  & \\
     \text{K$^{\tmop{it}}_{\tmop{gs}}$-machine } & \sigma_0 & \rightsquigarrow
     & \cdot \cdot \cdot & \rightsquigarrow & \sigma_n & \rightsquigarrow &
     \sigma_{n + 1} & \rightsquigarrow & \cdot \cdot \cdot\\
     & \; \longdownarrow \diamond &  &  &  & \; \longdownarrow \diamond &  &
     \; \longdownarrow \diamond &  & \\
     \text{K$_{\tmop{gs}}$-machine } & \sigma^{\diamond}_0 & \rightsquigarrow
     & \cdot \cdot \cdot & \rightsquigarrow & \sigma^{\diamond}_n &
     \rightsquigarrow & \sigma^{\diamond}_{n + 1} & \rightsquigarrow & \cdot
     \cdot \cdot
   \end{array} \]
The composition of simulation $\left( \mathsf{-} \right)^{\diamond}$ and
$\left( \mathsf{-} \right)^{\star}$ gives us a sound and complete lock-step
simulation of the K$_{\tmop{gs}}$-machine by the K$_{\tmop{ct}}$-machine.

\section{Conclusion and future work}

We have defined and formally proved the correctness of an abstract machine
which provides a direct computational interpretation of the Constant Domain
logic. However, as mentioned in the introduction, this work is a stepping
stone towards a computational interpretation of duality in intuitionistic
logic. Starting from the reduction semantics of proof terms of
bi-intuitionistic logic (subtractive logic) {\cite{Crolard04}}, it should be
possible to extend the coroutine machine to account for first-class
coroutines.

These future results should then be compared with other related works, such as
Curien and Herbelin's pioneering article on the duality of computation
{\cite{Curien00}}, or more recently, Bellin and Menti's work on the
$\pi$-calculus and co-intuitionistic logic {\cite{Bellin14}}, Kimura and
Tatsuta's Dual Calculus {\cite{Kimura09}} and Eades, Stump and McCleeary's
Dualized simple type theory {\cite{Eades16}}.

\paragraph*{Acknowledgments.}I would like to thank Nuria Brede for numerous
discussions on the ``safe'' $\lambda \mu$-calculus and useful comments on
earlier versions of this work. I am also very grateful to Olivier Danvy and
Ugo de'Liguoro for giving me the opportunity to present these results at WoC
2015, and for their work as editors of this special issue. Finally, I would
like to thank anonymous referees for their very constructive comments, which
helped me to substantially improve the manuscript.

\bibliographystyle{eptcs}\bibliography{biblio}

\begin{thebibliography}{10}
\providecommand{\bibitemdeclare}[2]{}
\providecommand{\surnamestart}{}
\providecommand{\surnameend}{}
\providecommand{\urlprefix}{Available at }
\providecommand{\url}[1]{\texttt{#1}}
\providecommand{\href}[2]{\texttt{#2}}
\providecommand{\urlalt}[2]{\href{#1}{#2}}
\providecommand{\doi}[1]{doi:\urlalt{http://dx.doi.org/#1}{#1}}
\providecommand{\bibinfo}[2]{#2}

\bibitemdeclare{article}{Anderson76}
\bibitem{Anderson76}
\bibinfo{author}{B.~\surnamestart Anderson\surnameend} (\bibinfo{year}{1976}):
  \emph{\bibinfo{title}{The Samefringe Problem}}.
\newblock {\sl \bibinfo{journal}{SIGART Bull.}} \bibinfo{volume}{60}, pp.
  \bibinfo{pages}{4--4}.

\bibitemdeclare{inproceedings}{Anton10}
\bibitem{Anton10}
\bibinfo{author}{K.~\surnamestart Anton\surnameend} \&
  \bibinfo{author}{P.~\surnamestart Thiemann\surnameend}
  (\bibinfo{year}{2010}): \emph{\bibinfo{title}{Towards Deriving Type Systems
  and Implementations for Coroutines}}.
\newblock In \bibinfo{editor}{Kazunori \surnamestart Ueda\surnameend}, editor:
  {\sl \bibinfo{booktitle}{Programming Languages and Systems -- 8th Asian
  Symposium, APLAS 2010}}, {\sl \bibinfo{series}{LNCS}} \bibinfo{volume}{6461},
  \bibinfo{publisher}{Springer}, \bibinfo{address}{Shanghai, China}, pp.
  \bibinfo{pages}{63--79}, \doi{10.1007/ 978-3-642-17164-2\_6}.

\bibitemdeclare{inproceedings}{Berardi94b}
\bibitem{Berardi94b}
\bibinfo{author}{F.~\surnamestart Barbanera\surnameend} \&
  \bibinfo{author}{S.~\surnamestart Berardi\surnameend} (\bibinfo{year}{1994}):
  \emph{\bibinfo{title}{{Extracting Constructive Content from Classical Logic
  via Control-like Reductions}}}.
\newblock In: {\sl \bibinfo{booktitle}{LNCS}}, \bibinfo{volume}{662},
  \bibinfo{publisher}{Springer-Verlag}, pp. \bibinfo{pages}{47--59},
  \doi{10.1.1.120.386}.

\bibitemdeclare{article}{Bellin14}
\bibitem{Bellin14}
\bibinfo{author}{G.~\surnamestart Bellin\surnameend} \&
  \bibinfo{author}{A.~\surnamestart Menti\surnameend} (\bibinfo{year}{2014}):
  \emph{\bibinfo{title}{{On the $\pi$-calculus and Co-intuitionistic Logic.
  Notes on Logic for Concurrency and $\lambda$P Systems}}}.
\newblock {\sl \bibinfo{journal}{Fundamenta Informaticae}}
  \bibinfo{volume}{130}(\bibinfo{number}{1}), pp. \bibinfo{pages}{21--65},
  \doi{10.3233/FI-2014-981}.

\bibitemdeclare{article}{Biernacka07}
\bibitem{Biernacka07}
\bibinfo{author}{M.~\surnamestart Biernacka\surnameend} \&
  \bibinfo{author}{O.~\surnamestart Danvy\surnameend} (\bibinfo{year}{2007}):
  \emph{\bibinfo{title}{{A Syntactic Correspondence between Context-Sensitive
  Calculi and Abstract Machines}}}.
\newblock {\sl \bibinfo{journal}{Theoretical Computer Science}}
  \bibinfo{volume}{375}, \doi{10.1016/j.tcs.2006.12.028}.

\bibitemdeclare{article}{Biernacki06}
\bibitem{Biernacki06}
\bibinfo{author}{D.~\surnamestart Biernacki\surnameend},
  \bibinfo{author}{O.~\surnamestart Danvy\surnameend} \&
  \bibinfo{author}{C.~\surnamestart Shan\surnameend} (\bibinfo{year}{2006}):
  \emph{\bibinfo{title}{On the Static and Dynamic Extents of Delimited
  Continuations}}.
\newblock {\sl \bibinfo{journal}{Science of Computer Programming}}
  \bibinfo{volume}{60}(\bibinfo{number}{3}), pp. \bibinfo{pages}{274--297},
  \doi{10.1016/j.scico.2006.01.002}.

\bibitemdeclare{mastersthesis}{Brede09}
\bibitem{Brede09}
\bibinfo{author}{N.~\surnamestart Brede\surnameend} (\bibinfo{year}{2009}):
  \emph{\bibinfo{title}{{$\lambda\mu$PRL - A Proof Refinement Calculus for
  Classical Reasoning in Computational Type Theory}}}.
\newblock Master's thesis, \bibinfo{school}{University of Potsdam}.
\newblock \urlprefix\url{http://www.cs.uni-potsdam.de/~brede}.

\bibitemdeclare{article}{Clint73}
\bibitem{Clint73}
\bibinfo{author}{M.~\surnamestart Clint\surnameend} (\bibinfo{year}{1973}):
  \emph{\bibinfo{title}{{Program proving: Coroutines}}}.
\newblock {\sl \bibinfo{journal}{Acta Informatica}}
  \bibinfo{volume}{2}(\bibinfo{number}{1}), pp. \bibinfo{pages}{50--63},
  \doi{10.1007/BF00571463}.

\bibitemdeclare{article}{Conway63}
\bibitem{Conway63}
\bibinfo{author}{M.~E. \surnamestart Conway\surnameend} (\bibinfo{year}{1963}):
  \emph{\bibinfo{title}{{Design of a separable transition-diagram compiler}}}.
\newblock {\sl \bibinfo{journal}{Commun. ACM}}
  \bibinfo{volume}{6}(\bibinfo{number}{7}), pp. \bibinfo{pages}{396--408},
  \doi{10.1145/366663.366704}.

\bibitemdeclare{misc}{Crolard96}
\bibitem{Crolard96}
\bibinfo{author}{T.~\surnamestart Crolard\surnameend} (\bibinfo{year}{1996}):
  \emph{\bibinfo{title}{{Extension de l'Isomorphisme de {C}urry-{H}oward au
  Traitement des Exceptions (application d'une {\'e}tude de la dualit{\'e} en
  logique intuitionniste)}}}.
\newblock \bibinfo{howpublished}{Th{\`e}se de Doctorat. Universit{\'e} Paris
  7}.

\bibitemdeclare{article}{Crolard98a}
\bibitem{Crolard98a}
\bibinfo{author}{T.~\surnamestart Crolard\surnameend} (\bibinfo{year}{1999}):
  \emph{\bibinfo{title}{{A confluent lambda-calculus with a catch/throw
  mechanism}}}.
\newblock {\sl \bibinfo{journal}{Journal of Functional Programming}}
  \bibinfo{volume}{9}(\bibinfo{number}{6}), pp. \bibinfo{pages}{625--647},
  \doi{10.1017/S0956796899003512}.

\bibitemdeclare{article}{Crolard01}
\bibitem{Crolard01}
\bibinfo{author}{T.~\surnamestart Crolard\surnameend} (\bibinfo{year}{2001}):
  \emph{\bibinfo{title}{{S}ubtractive {L}ogic}}.
\newblock {\sl \bibinfo{journal}{Theoretical Computer Science}}
  \bibinfo{volume}{254}(\bibinfo{number}{1--2}), pp. \bibinfo{pages}{151--185},
  \doi{10.1016/S0304-3975(99)00124-3}.

\bibitemdeclare{article}{Crolard04}
\bibitem{Crolard04}
\bibinfo{author}{T.~\surnamestart Crolard\surnameend} (\bibinfo{year}{2004}):
  \emph{\bibinfo{title}{{A} {F}ormul{\ae}-as-{T}ypes {I}nterpretation of
  {S}ubtractive {L}ogic}}.
\newblock {\sl \bibinfo{journal}{Journal of Logic and Computation}}
  \bibinfo{volume}{14}(\bibinfo{number}{4}), pp. \bibinfo{pages}{529--570},
  \doi{10.1093/logcom/14.4.529}.

\bibitemdeclare{techreport}{Crolard15}
\bibitem{Crolard15}
\bibinfo{author}{T.~\surnamestart Crolard\surnameend} (\bibinfo{year}{2015}):
  \emph{\bibinfo{title}{{A verified abstract machine for functional coroutines
  - Coq formalization}}}.
\newblock \bibinfo{type}{Technical Report}, \bibinfo{institution}{CEDRIC -
  Conservatoire National des Arts et M{\'e}tiers}.
\newblock \urlprefix\url{http://cedric.cnam.fr/cpr/crolard/publications}.

\bibitemdeclare{inproceedings}{Curien00}
\bibitem{Curien00}
\bibinfo{author}{P.-L. \surnamestart Curien\surnameend} \&
  \bibinfo{author}{H.~\surnamestart Herbelin\surnameend}
  (\bibinfo{year}{2000}): \emph{\bibinfo{title}{{The duality of computation.}}}
\newblock In: {\sl \bibinfo{booktitle}{Proceedings of the {ACM} SIGPLAN
  International Conference on Functional Programming ({ICFP}'00)}},
  \bibinfo{publisher}{ACM Press}, \bibinfo{address}{New York, USA}, pp.
  \bibinfo{pages}{233--243}, \doi{10.1145/351240.351262}.

\bibitemdeclare{book}{Dahl72}
\bibitem{Dahl72}
\bibinfo{author}{O.~J. \surnamestart Dahl\surnameend}, \bibinfo{author}{E.~W.
  \surnamestart Dijkstra\surnameend} \& \bibinfo{author}{C.~A.~R. \surnamestart
  Hoare\surnameend} (\bibinfo{year}{1972}): \emph{\bibinfo{title}{{Structured
  programming}}}.
\newblock \bibinfo{publisher}{Academic Press}.

\bibitemdeclare{article}{Dahl66}
\bibitem{Dahl66}
\bibinfo{author}{O.-J. \surnamestart Dahl\surnameend} \&
  \bibinfo{author}{K.~\surnamestart Nygaard\surnameend} (\bibinfo{year}{1966}):
  \emph{\bibinfo{title}{{SIMULA: an ALGOL-based simulation language}}}.
\newblock {\sl \bibinfo{journal}{Commun. ACM}}
  \bibinfo{volume}{9}(\bibinfo{number}{9}), pp. \bibinfo{pages}{671--678},
  \doi{10.1145/365813.365819}.

\bibitemdeclare{inproceedings}{Danvy08}
\bibitem{Danvy08}
\bibinfo{author}{O.~\surnamestart Danvy\surnameend} (\bibinfo{year}{2008}):
  \emph{\bibinfo{title}{Defunctionalized Interpreters for Programming
  Languages}}.
\newblock In: {\sl \bibinfo{booktitle}{Proceedings of the {ACM} SIGPLAN
  International Conference on Functional Programming (ICFP'08)}},
  \bibinfo{publisher}{ACM Press}, \bibinfo{address}{New York, USA}, pp.
  \bibinfo{pages}{131--142}, \doi{10.1145/1411204.1411206}.

\bibitemdeclare{article}{Danvy07}
\bibitem{Danvy07}
\bibinfo{author}{O.~\surnamestart {Danvy, editor}\surnameend}
  (\bibinfo{year}{2007}): \emph{\bibinfo{title}{{Special Issue on the Krivine
  Machine}}}.
\newblock {\sl \bibinfo{journal}{Higher-Order and Symbolic Computation}}
  \bibinfo{volume}{20}(\bibinfo{number}{3}), \doi{10.1007/s10990-007-9021-1}.

\bibitemdeclare{article}{Dybvig89}
\bibitem{Dybvig89}
\bibinfo{author}{R.~K. \surnamestart Dybvig\surnameend} \&
  \bibinfo{author}{R.~\surnamestart Hieb\surnameend} (\bibinfo{year}{1989}):
  \emph{\bibinfo{title}{{Engines From Continuations}}}.
\newblock {\sl \bibinfo{journal}{Comput. Lang}}
  \bibinfo{volume}{14}(\bibinfo{number}{2}), pp. \bibinfo{pages}{109--123},
  \doi{10.1016/0096-0551(89)90018-0}.

\bibitemdeclare{article}{Eades16}
\bibitem{Eades16}
\bibinfo{author}{H.~\surnamestart Eades\surnameend},
  \bibinfo{author}{A.~\surnamestart Stump\surnameend} \&
  \bibinfo{author}{R.~\surnamestart McCleeary\surnameend}
  (\bibinfo{year}{2016}): \emph{\bibinfo{title}{Dualized simple type theory}}.
\newblock {\sl \bibinfo{journal}{Submitted to Logical Methods in Computer
  Science}}.

\bibitemdeclare{article}{Fortune83}
\bibitem{Fortune83}
\bibinfo{author}{S.~\surnamestart Fortune\surnameend},
  \bibinfo{author}{D.~\surnamestart Leivant\surnameend} \&
  \bibinfo{author}{M.~\surnamestart O'Donnell\surnameend}
  (\bibinfo{year}{1983}): \emph{\bibinfo{title}{The Expressiveness of Simple
  and Second-Order Type Structures}}.
\newblock {\sl \bibinfo{journal}{J. ACM}}
  \bibinfo{volume}{30}(\bibinfo{number}{1}), pp. \bibinfo{pages}{151--185},
  \doi{10.1145/322358.322370}.

\bibitemdeclare{article}{Wand86}
\bibitem{Wand86}
\bibinfo{author}{D.~P. \surnamestart Friedman\surnameend},
  \bibinfo{author}{C.~T. \surnamestart Haynes\surnameend} \&
  \bibinfo{author}{M.~\surnamestart Wand\surnameend} (\bibinfo{year}{1986}):
  \emph{\bibinfo{title}{{Obtaining Coroutines with Continuations}}}.
\newblock {\sl \bibinfo{journal}{Journal of Computer Languages}}
  \bibinfo{volume}{11}(\bibinfo{number}{3/4}), pp. \bibinfo{pages}{143--153},
  \doi{10.1016/0096-0551(86)90007-X}.

\bibitemdeclare{article}{Gore08}
\bibitem{Gore08}
\bibinfo{author}{R.~\surnamestart Gor{\'e}\surnameend} \&
  \bibinfo{author}{L.~\surnamestart Postniece\surnameend}
  (\bibinfo{year}{2010}): \emph{\bibinfo{title}{{Combining derivations and
  refutations for cut-free completeness in bi-intuitionistic logic}}}.
\newblock {\sl \bibinfo{journal}{Journal of Logic and Computation}},
  \doi{10.1093/logcom/exn067}.

\bibitemdeclare{article}{Gornemann71}
\bibitem{Gornemann71}
\bibinfo{author}{S.~\surnamestart G{\"o}rnemann\surnameend}
  (\bibinfo{year}{1971}): \emph{\bibinfo{title}{{A logic stronger than
  intuitionism}}}.
\newblock {\sl \bibinfo{journal}{The Journal of Symbolic Logic}}
  \bibinfo{volume}{36}, pp. \bibinfo{pages}{249--261}, \doi{10.2307/2270260}.

\bibitemdeclare{article}{Greussay76}
\bibitem{Greussay76}
\bibinfo{author}{P.~\surnamestart Greussay\surnameend} (\bibinfo{year}{1976}):
  \emph{\bibinfo{title}{An Iterative Lisp Solution to the Samefringe Problem}}.
\newblock {\sl \bibinfo{journal}{SIGART Bull.}} \bibinfo{volume}{59}, pp.
  \bibinfo{pages}{14--14}, \doi{10.1145/1045270.1045273}.

\bibitemdeclare{inproceedings}{Griffin90}
\bibitem{Griffin90}
\bibinfo{author}{T.~G. \surnamestart Griffin\surnameend}
  (\bibinfo{year}{1990}): \emph{\bibinfo{title}{{A formul{\ae}-as-types notion
  of control}}}.
\newblock In: {\sl \bibinfo{booktitle}{Conference Record of the 17th Annual ACM
  Symposium on Principles of Programming Langages}}, pp.
  \bibinfo{pages}{47--58}, \doi{10.1145/96709.96714}.

\bibitemdeclare{article}{deGroote98}
\bibitem{deGroote98}
\bibinfo{author}{P.~\surnamestart de~Groote\surnameend} (\bibinfo{year}{1998}):
  \emph{\bibinfo{title}{{An environment machine for the lambda-mu-calculus.}}}
\newblock {\sl \bibinfo{journal}{Mathematical Structure in Computer Science}}
  \bibinfo{volume}{8}, pp. \bibinfo{pages}{637--669},
  \doi{10.1017/S0960129598002667}.

\bibitemdeclare{incollection}{deGroote01}
\bibitem{deGroote01}
\bibinfo{author}{P.~\surnamestart de~Groote\surnameend} (\bibinfo{year}{2001}):
  \emph{\bibinfo{title}{Strong Normalization of Classical Natural Deduction
  with Disjunction}}.
\newblock In \bibinfo{editor}{S.~\surnamestart Abramsky\surnameend}, editor:
  {\sl \bibinfo{booktitle}{Typed Lambda Calculi and Applications}}, {\sl
  \bibinfo{series}{LNCS}} \bibinfo{volume}{2044},
  \bibinfo{publisher}{Springer}, pp. \bibinfo{pages}{182--196},
  \doi{10.1007/3-540-45413-6\_17}.

\bibitemdeclare{article}{Grzegorczyk64}
\bibitem{Grzegorczyk64}
\bibinfo{author}{A.~\surnamestart Grzegorczyk\surnameend}
  (\bibinfo{year}{1964}): \emph{\bibinfo{title}{{A philosophically plausible
  formal interpretation of intuitionistic logic.}}}
\newblock {\sl \bibinfo{journal}{Nederl. Akad. Wet., Proc., Ser. A}}
  \bibinfo{volume}{67}, pp. \bibinfo{pages}{596--601}, \doi{10.2307/2271877}.

\bibitemdeclare{article}{MacQueen93}
\bibitem{MacQueen93}
\bibinfo{author}{R.~\surnamestart Harper\surnameend}, \bibinfo{author}{B.~F.
  \surnamestart Duba\surnameend} \& \bibinfo{author}{D.~\surnamestart
  MacQueen\surnameend} (\bibinfo{year}{1993}): \emph{\bibinfo{title}{Typing
  first-class continuations in {ML}}}.
\newblock {\sl \bibinfo{journal}{Journal of Functional Programming}}
  \bibinfo{volume}{3}(\bibinfo{number}{4}), pp. \bibinfo{pages}{465--484},
  \doi{10.1017/S095679680000085X}.

\bibitemdeclare{techreport}{Kashima91}
\bibitem{Kashima91}
\bibinfo{author}{R.~\surnamestart Kashima\surnameend} (\bibinfo{year}{1991}):
  \emph{\bibinfo{title}{{Cut-Elimination for the intermediate logic CD}}}.
\newblock \bibinfo{type}{Research Report on Information Sciences}
  \bibinfo{number}{C100}, \bibinfo{institution}{Institute of Technology},
  \bibinfo{address}{Tokyo}.

\bibitemdeclare{inproceedings}{Kimura09}
\bibitem{Kimura09}
\bibinfo{author}{D.~\surnamestart Kimura\surnameend} \&
  \bibinfo{author}{M.~\surnamestart Tatsuta\surnameend} (\bibinfo{year}{2009}):
  \emph{\bibinfo{title}{Dual Calculus with Inductive and Coinductive Types}}.
\newblock In \bibinfo{editor}{Ralf \surnamestart Treinen\surnameend}, editor:
  {\sl \bibinfo{booktitle}{Rewriting Techniques and Applications, 20th
  International Conference, RTA 2009, Bras\'{\i}lia, Brazil}}, {\sl
  \bibinfo{series}{LNCS}} \bibinfo{volume}{5595},
  \bibinfo{publisher}{Springer}, pp. \bibinfo{pages}{224--238},
  \doi{10.1007/978-3-642-02348-4\_16}.

\bibitemdeclare{book}{Knuth97}
\bibitem{Knuth97}
\bibinfo{author}{D.~E. \surnamestart Knuth\surnameend} (\bibinfo{year}{1997}):
  \emph{\bibinfo{title}{{The Art of Computer Programming, Volume I: Fundamental
  Algorithms}}}, \bibinfo{edition}{3rd edition} edition.
\newblock \bibinfo{publisher}{Addison-Wesley}.

\bibitemdeclare{article}{Krivine94}
\bibitem{Krivine94}
\bibinfo{author}{J.-L. \surnamestart Krivine\surnameend}
  (\bibinfo{year}{1994}): \emph{\bibinfo{title}{{Classical logic, storage
  operators and second order $\lambda$-calculus}}}.
\newblock {\sl \bibinfo{journal}{Ann. of Pure and Appl. Logic}}
  \bibinfo{volume}{68}, pp. \bibinfo{pages}{53--78},
  \doi{10.1016/0168-0072(94)90047-7}.

\bibitemdeclare{article}{Leivant02}
\bibitem{Leivant02}
\bibinfo{author}{D.~\surnamestart Leivant\surnameend} (\bibinfo{year}{2002}):
  \emph{\bibinfo{title}{{Intrinsic reasoning about functional programs {I}:
  first order theories}}}.
\newblock {\sl \bibinfo{journal}{Annals of Pure and Applied Logic}}
  \bibinfo{volume}{114}(\bibinfo{number}{1-3}), pp. \bibinfo{pages}{117--153},
  \doi{10.1016/S0168-0072(01)00078-1}.

\bibitemdeclare{article}{Lopez-Escobar83}
\bibitem{Lopez-Escobar83}
\bibinfo{author}{E.~G.~K. \surnamestart Lopez-Escobar\surnameend}
  (\bibinfo{year}{1983}): \emph{\bibinfo{title}{{A Second Paper ``On the
  Interpolation Theorem for the Logic of Constant Domains"}}}.
\newblock {\sl \bibinfo{journal}{The Journal of Symbolic Logic}}
  \bibinfo{volume}{48}(\bibinfo{number}{3}), pp. \bibinfo{pages}{595--599},
  \doi{10.2307/2273451}.
\newblock \urlprefix\url{http://www.jstor.org/stable/2273451}.

\bibitemdeclare{book}{Marlin80}
\bibitem{Marlin80}
\bibinfo{author}{C.~D. \surnamestart Marlin\surnameend} (\bibinfo{year}{1980}):
  \emph{\bibinfo{title}{{Coroutines: A Programming Methodology, a Language
  Design and an Implementation}}}.
\newblock \bibinfo{publisher}{Springer-Verlag New York, Inc.},
  \bibinfo{address}{Secaucus, NJ, USA}, \doi{10.1007/3-540-10256-6}.

\bibitemdeclare{article}{McCarthy77}
\bibitem{McCarthy77}
\bibinfo{author}{J.~\surnamestart McCarthy\surnameend} (\bibinfo{year}{1977}):
  \emph{\bibinfo{title}{Another SAMEFRINGE}}.
\newblock {\sl \bibinfo{journal}{SIGART Bull.}} \bibinfo{volume}{61}, pp.
  \bibinfo{pages}{4--4}.

\bibitemdeclare{article}{Mints13}
\bibitem{Mints13}
\bibinfo{author}{G.~\surnamestart Mints\surnameend},
  \bibinfo{author}{G.~\surnamestart Olkhovikov\surnameend} \&
  \bibinfo{author}{A.~\surnamestart Urquhart\surnameend}
  (\bibinfo{year}{2013}): \emph{\bibinfo{title}{Failure of Interpolation in
  Constant Domain Intuitionistic Logic}}.
\newblock {\sl \bibinfo{journal}{The Journal of Symbolic Logic}}
  \bibinfo{volume}{78}, pp. \bibinfo{pages}{937--950},
  \doi{10.2178/jsl.7803120}.
\newblock
  \urlprefix\url{http://journals.cambridge.org/article_S0022481200126672}.

\bibitemdeclare{techreport}{Moura04a}
\bibitem{Moura04a}
\bibinfo{author}{A.~L. \surnamestart de~Moura\surnameend} \&
  \bibinfo{author}{R.~\surnamestart Ierusalimschy\surnameend}
  (\bibinfo{year}{2004}): \emph{\bibinfo{title}{{Revisiting Coroutines}}}.
\newblock \bibinfo{type}{MCC} \bibinfo{number}{15/04},
  \bibinfo{institution}{PUC-Rio}, \bibinfo{address}{Rio de Janeiro, RJ},
  \doi{10.1145/1462166.1462167}.

\bibitemdeclare{article}{Moura04}
\bibitem{Moura04}
\bibinfo{author}{A.~L. \surnamestart de~Moura\surnameend},
  \bibinfo{author}{N.~\surnamestart Rodriguez\surnameend} \&
  \bibinfo{author}{R.~\surnamestart Ierusalimschy\surnameend}
  (\bibinfo{year}{2004}): \emph{\bibinfo{title}{{Coroutines in Lua}}}.
\newblock {\sl \bibinfo{journal}{Journal of Universal Computer Science}}
  \bibinfo{volume}{10}(\bibinfo{number}{7}), pp. \bibinfo{pages}{910--925},
  \doi{10.3217/jucs-010-07-0910}.

\bibitemdeclare{techreport}{Murthy91b}
\bibitem{Murthy91b}
\bibinfo{author}{C.~R. \surnamestart Murthy\surnameend} (\bibinfo{year}{1991}):
  \emph{\bibinfo{title}{{Classical proofs as programs: How, when, and why}}}.
\newblock \bibinfo{type}{Technical Report} \bibinfo{number}{91-1215},
  \bibinfo{institution}{Cornell University, Department of Computer Science}.

\bibitemdeclare{inproceedings}{Parigot94}
\bibitem{Parigot94}
\bibinfo{author}{M.~\surnamestart Parigot\surnameend} (\bibinfo{year}{1993}):
  \emph{\bibinfo{title}{{Strong normalization for second order classical
  natural deduction}}}.
\newblock In: {\sl \bibinfo{booktitle}{Proceedings of the eighth annual {IEEE}
  symposium on logic in computer science}}, pp. \bibinfo{pages}{39--46},
  \doi{10.1109/LICS.1993.287602}.

\bibitemdeclare{article}{Pinto09}
\bibitem{Pinto09}
\bibinfo{author}{L.~\surnamestart Pinto\surnameend} \&
  \bibinfo{author}{T.~\surnamestart Uustalu\surnameend} (\bibinfo{year}{2009}):
  \emph{\bibinfo{title}{{Proof Search and Counter-Model Construction for
  Bi-intuitionistic Propositional Logic with Labelled Sequents}}}.
\newblock {\sl \bibinfo{journal}{Automated Reasoning with Analytic Tableaux and
  Related Methods}}, pp. \bibinfo{pages}{295--309},
  \doi{10.1007/978-3-642-02716-1\_22}.

\bibitemdeclare{inproceedings}{Pinto10}
\bibitem{Pinto10}
\bibinfo{author}{L.~\surnamestart Pinto\surnameend} \&
  \bibinfo{author}{T.~\surnamestart Uustalu\surnameend} (\bibinfo{year}{2010}):
  \emph{\bibinfo{title}{{Relating Sequent Calculi for Bi-intuitionistic
  Propositional Logic}}}.
\newblock In \bibinfo{editor}{S.~\surnamestart van Bakel\surnameend},
  \bibinfo{editor}{S.~\surnamestart Berardi\surnameend} \&
  \bibinfo{editor}{U.~\surnamestart Berger\surnameend}, editors: {\sl
  \bibinfo{booktitle}{Proc. of the 3rd Workshop on Classical logic and
  Computation}}, \bibinfo{organization}{Masarykova Univ.}, pp.
  \bibinfo{pages}{68--85}, \doi{10.4204/EPTCS.47.7}.

\bibitemdeclare{article}{Prenner71}
\bibitem{Prenner71}
\bibinfo{author}{C.~J. \surnamestart Prenner\surnameend}
  (\bibinfo{year}{1971}): \emph{\bibinfo{title}{The Control Structure
  Facilities of ECL}}.
\newblock {\sl \bibinfo{journal}{SIGPLAN Not.}}
  \bibinfo{volume}{6}(\bibinfo{number}{12}), pp. \bibinfo{pages}{104--112},
  \doi{10.1145/800006.807990}.

\bibitemdeclare{article}{Pym00}
\bibitem{Pym00}
\bibinfo{author}{D.~\surnamestart Pym\surnameend},
  \bibinfo{author}{E.~\surnamestart Ritter\surnameend} \&
  \bibinfo{author}{L.~\surnamestart Wallen\surnameend} (\bibinfo{year}{2000}):
  \emph{\bibinfo{title}{{On the intuitionistic force of classical search}}}.
\newblock {\sl \bibinfo{journal}{Theoretical Computer Science}}
  \bibinfo{volume}{232(1-2)}, pp. \bibinfo{pages}{299--333},
  \doi{10.1016/S0304-3975(99)00178-4}.

\bibitemdeclare{inproceedings}{Rauszer74}
\bibitem{Rauszer74}
\bibinfo{author}{C.~\surnamestart Rauszer\surnameend} (\bibinfo{year}{1974}):
  \emph{\bibinfo{title}{{Semi-Boolean algebras and their applications to
  intuitionistic logic with dual operations}}}.
\newblock In: {\sl \bibinfo{booktitle}{Fundamenta Mathematicae}},
  \bibinfo{volume}{83}, pp. \bibinfo{pages}{219--249}.

\bibitemdeclare{incollection}{Rauszer80}
\bibitem{Rauszer80}
\bibinfo{author}{C.~\surnamestart Rauszer\surnameend} (\bibinfo{year}{1980}):
  \emph{\bibinfo{title}{{An algebraic and {K}ripke-style approach to a certain
  extension of intuitionistic logic}}}.
\newblock In: {\sl \bibinfo{booktitle}{Dissertationes Mathematicae}},
  \bibinfo{volume}{167}, \bibinfo{publisher}{Institut Math{\'e}matique de
  l'Acad{\'e}mie Polonaise des Sciences}, pp. \bibinfo{pages}{1--67}.

\bibitemdeclare{inproceedings}{Rehof94}
\bibitem{Rehof94}
\bibinfo{author}{N.~J. \surnamestart Rehof\surnameend} \&
  \bibinfo{author}{M.~H. \surnamestart S{\o}rensen\surnameend}
  (\bibinfo{year}{1994}): \emph{\bibinfo{title}{{The
  $\lambda_\Delta$-calculus}}}.
\newblock In: {\sl \bibinfo{booktitle}{Theoretical Aspects of Computer
  Software}}, {\sl \bibinfo{series}{LNCS}} \bibinfo{volume}{542},
  \bibinfo{publisher}{Springer-Verlag}, pp. \bibinfo{pages}{516--542},
  \doi{10.1007/3-540-57887-0\_113}.

\bibitemdeclare{inproceedings}{Reppy95}
\bibitem{Reppy95}
\bibinfo{author}{J.~H. \surnamestart Reppy\surnameend} (\bibinfo{year}{1995}):
  \emph{\bibinfo{title}{First-class Synchronous Operations}}.
\newblock In: {\sl \bibinfo{booktitle}{Proceedings of the International
  Workshop on Theory and Practice of Parallel Programming}}, {\sl
  \bibinfo{series}{LNCS}} \bibinfo{volume}{907},
  \bibinfo{publisher}{Springer-Verlag}, \bibinfo{address}{London, UK}, pp.
  \bibinfo{pages}{235--252}, \doi{10.1007/BFb0026573}.

\bibitemdeclare{article}{Shimura94}
\bibitem{Shimura94}
\bibinfo{author}{T.~\surnamestart Shimura\surnameend} \&
  \bibinfo{author}{R.~\surnamestart Kashima\surnameend} (\bibinfo{year}{1994}):
  \emph{\bibinfo{title}{{Cut-Elimination Theorem for the Logic of Constant
  Domains}}}.
\newblock {\sl \bibinfo{journal}{Math. Log. Q}} \bibinfo{volume}{40}, pp.
  \bibinfo{pages}{153--172}, \doi{10.1002/malq.19940400203}.

\bibitemdeclare{techreport}{Strachey74}
\bibitem{Strachey74}
\bibinfo{author}{C.~\surnamestart Strachey\surnameend} \&
  \bibinfo{author}{C.~P. \surnamestart Wadsworth\surnameend}
  (\bibinfo{year}{1974}): \emph{\bibinfo{title}{Continuations: A Mathematical
  Semantics for Handling Full Jumps}}.
\newblock \bibinfo{type}{Technical Monograph} \bibinfo{number}{PRG-11},
  \bibinfo{institution}{Oxford University Computing Laboratory, Programming
  Research Group}, \bibinfo{address}{Oxford, England}.
\newblock \bibinfo{note}{Reprinted in Higher-Order and Symbolic Computation
  13(1/2):135--152, 2000, with a foreword~\cite{Wadsworth00}}.

\bibitemdeclare{article}{Streicher98}
\bibitem{Streicher98}
\bibinfo{author}{T.~\surnamestart Streicher\surnameend} \&
  \bibinfo{author}{B.~\surnamestart Reus\surnameend} (\bibinfo{year}{1998}):
  \emph{\bibinfo{title}{{Classical Logic, Continuation Semantics and Abstract
  Machines}}}.
\newblock {\sl \bibinfo{journal}{Journal of Functional Programming}}
  \bibinfo{volume}{8}(\bibinfo{number}{6}), pp. \bibinfo{pages}{543--572},
  \doi{10.1017/S0956796898003141}.

\bibitemdeclare{misc}{Open97}
\bibitem{Open97}
\bibinfo{author}{\surnamestart {The Open Group}\surnameend}
  (\bibinfo{year}{1997}): \emph{\bibinfo{title}{{The {S}ingle {UNIX}
  {S}pecification, {V}ersion 2}}}.
\newblock \urlprefix\url{http://www.UNIX-systems.org/online.html}.

\bibitemdeclare{book}{Troelstra73}
\bibitem{Troelstra73}
\bibinfo{author}{A.~S. \surnamestart Troelstra\surnameend}
  (\bibinfo{year}{1973}): \emph{\bibinfo{title}{{Metamathematical Investigation
  of Intuitionistic Arithmetic and Analysis}}}.
\newblock {\sl \bibinfo{series}{Lecture Notes in Mathematics}}
  \bibinfo{volume}{344}, \bibinfo{publisher}{Springer-Verlag},
  \bibinfo{address}{Berlin}, \doi{10.1007/BFb0066742}.

\bibitemdeclare{article}{Wadsworth00}
\bibitem{Wadsworth00}
\bibinfo{author}{C.~P. \surnamestart Wadsworth\surnameend}
  (\bibinfo{year}{2000}): \emph{\bibinfo{title}{Continuations revisited}}.
\newblock {\sl \bibinfo{journal}{Higher-Order and Symbolic Computation}}
  \bibinfo{volume}{13}(\bibinfo{number}{1/2}), pp. \bibinfo{pages}{131--133},
  \doi{10.1023/A:1010074329461}.

\bibitemdeclare{book}{Wirth80}
\bibitem{Wirth80}
\bibinfo{author}{N.~\surnamestart Wirth\surnameend} \&
  \bibinfo{author}{J.~\surnamestart Mincer-Daszkiewicz\surnameend}
  (\bibinfo{year}{1980}): \emph{\bibinfo{title}{{Modula-2}}}.
\newblock \bibinfo{publisher}{ETH Zurich, Schweiz},
  \doi{10.3929/ethz-a-000189918}.

\end{thebibliography}

\end{document}